\documentclass[twoside]{book}   

\usepackage{latexsym,amssymb,rotate,epsf,crc1-2e,epsfig,amsfonts,amstext}
\usepackage{curves,dsfont,rotating,multicol,fancyheadings,graphicx,epic,eepic} % \usepackage{cite}
\usepackage{diagrams}
\usepackage{verbatim}
\usepackage{color}   
\newarrow{Is}=====
\newarrow{Mo}|---> 

\newcommand{\n}{\noindent}

\begin{document} 

\frontmatter
\pagenumbering{roman}

\tableofcontents

\mainmatter

\def\bR{\begin{color}{red}}
\def\bB{\begin{color}{blue}}
\def\bM{\begin{color}{magenta}}   
\def\bC{\begin{color}{cyan}}
\def\bW{\begin{color}{white}}
\def\bBl{\begin{color}{black}}
\def\bG{\begin{color}{green}}
\def\bY{\begin{color}{yellow}}
\def\e{\end{color}}

\newcommand{\bit}{\begin{itemize}}
\newcommand{\eit}{\end{itemize}\par\noindent} 
\newcommand{\ben}{\begin{enumerate}}
\newcommand{\een}{\end{enumerate}\par\noindent}
\newcommand{\beq}{\begin{equation}}
\newcommand{\eeq}{\end{equation}\par\noindent}
\newcommand{\beqa}{\begin{eqnarray*}}
\newcommand{\eeqa}{\end{eqnarray*}\par\noindent}
\newcommand{\beqn}{\begin{eqnarray}}
\newcommand{\eeqn}{\end{eqnarray}\par\noindent}

%\newarrow{Is}=====
%\newarrow{Mo}|---> 

\def\PP{{\rm P}} 
\def\HH{\mathcal{H}}
\def\CC{{\bf C}}
\def\II{{\rm I}}
\newcommand{\dd}{\llcorner}
\newcommand{\sdot}{\bullet}
\newcommand{\ddd}{\lrcorner}
\newcommand{\uu}{\ulcorner}
\newcommand{\uuu}{\urcorner}
\newcommand{\dist}{\mbox{\sf\footnotesize DIST}}

\chapter{Quantum measurements without sums}

\contributor{Bob Coecke and Dusko Pavlovic}

%\noindent Oxford University Computing Laboratory,\\ Wolfson Building, Parks Road,\\ Oxford OX1 3QD, %UK.\\ coecke@comlab.ox.ac.uk\\ \ \\ Kestrel Institute,\\ 3260 Hillview Avenue,\\ Palo Alto CA %94304, US.\\ dusko@kestrel.edu

\begin{abstract}
Sums play a prominent role in the formalisms of quantum mechanics, be it for mixing and superposing states, or for composing state spaces. Surprisingly,  a conceptual analysis of quantum measurement seems to suggest that quantum mechanics can be done without direct sums, \em expressed entirely in terms of the tensor product\em. The corresponding axioms define classical spaces as objects that allow copying and deleting data.  Indeed, the information exchange between the quantum and the classical worlds is essentially determined by their \em distinct capabilities to copy and delete data\em. The sums turn out to be an implicit implementation of this capability.   Realizing it through explicit axioms not only dispenses with the unnecessary structural baggage, but also allows a simple and intuitive \em graphical calculus\em.  
In category-theoretic terms, classical data types are  \em $\dagger$-compact Frobenius algebras\em, and quantum spectra underlying quantum measurements are \em Eilenberg-Moore coalgebras \em induced by these Frobenius algebras.
\end{abstract}

%\tableofcontents 

\section{Introduction} 

Ever since John von Neumann denounced, back in 1935 \cite{Redei}, his own foundation of quantum mechanics in terms of Hilbert spaces, there has been an ongoing search for a high-level, fully abstract formalism of quantum mechanics.  With the emergence of quantum information technology, this quest became more important than ever.  The low-level matrix manipulations in quantum informatics are akin to machine programming with bit strings from the early days of computing, which are of course  inadequate.\footnote{But while computing devices do manipulate strings of 0s and 1s, and high-level modern programming is a matter of providing a convenient interface with that process, the language for quantum information and computation we seek is not a convenient superstructure, but the meaningful infrastructure.}

A recent research thread, initiated by Abramsky and the first author \cite{AC1}, aims at recasting the quantum mechanical formalism in  {\em categorical\/} terms.  The upshot of categorical semantics is that it displays concepts in a \em compositional \em and \em typed \em framework. In the case of quantum mechanics, it uncovers the \em quantum information-flows \em \cite{LN2} which are hidden in the usual formalism.  Moreover, while the investigations of quantum structures have so far been predominantly academic, categorical semantics open an alley towards a practical, low-overhead tool for the design and analysis of quantum informatic protocols, versatile enough to capture both quantitative and qualitative aspects of quantum information  \cite{AC1,deLL,Paquette,CPP,Ross,Selinger2}.  In fact, some otherwise complicated quantum informatic protocols become trivial exercises in this framework \cite{Kindergarten}.  On the other hand, compared with the order-theoretic framework for quantum mechanics in terms of  Birkhoff-von Neumann's quantum logic \cite{Piron}, this categorical setting comes with logical derivations, topologically embodied into something as simple as ``yanking a rope''.\footnote{A closely related knot-theoretical scheme has been put forward by Kauffman in \cite{Kauffman}.}  Moreover, in terms of deductive mechanism, it turns out to be some kind of ``hyper-logic'' \cite{Ross}, as compared to Birkhoff-von Neumann logic  which as a consequence of being non-distributive fails to admit a deduction mechanism.  
 
The core of categorical semantics are \em $\dagger$-compact categories\em, originally proposed  in  \cite{AC1,AC1.5} under the name \em strongly compact closed categories\em, extending the structure of \em compact closed categories\em, which have been familiar in various communities since the 1970es \cite{KellyLaplaza}.   A salient 
%but at the same time very attractive --- "salient" ~ "attractive, convenient"
feature of \em categorical tensor calculi \em of this kind is that they admit sound and complete graphical representations, in the sense that \em a well-typed equation in such a tensor calculus is provable from its axioms if and only if the graphical interpretation of that equation is valid in the graphical language\em. Various graphical calculi have been an important vehicle of computation in physics \cite[and subsequent work]{Penrose}, and a prominent research topic of category theory e.g.~\cite{Kelly,KellyLaplaza,JS}. Soundness and completeness of the graphical language of $\dagger$-compact categories, which can be viewed as a two-dimensional formalization and extension of Dirac's bra-ket notation \cite{Kindergarten},  has been demonstrated by Selinger in \cite{Selinger2}. Besides this reference, the interested reader may wish to consult \cite{Abr,JS,Selinger2} for methods and proofs, and \cite{Kindergarten, Cats} and also Baez's \cite{Baez} for a more leisurely introduction into $\dagger$-categories. 

An important aspect of the $\dagger$-compact semantics of quantum protocols proposed in \cite{AC1,deLL,Selinger2} was the interplay of the multiplicative and additive structures of tensor products and direct sums, respectively. The direct sums (in fact biproducts, since all compact categories are self-dual) seemed essential for specifying classical data types, families of mutually orthogonal projectors, and ultimately for defining measurements.  The drawback of this was that the additive types do not yield to a simple graphical calculus; in fact, they make it unusable for many practical purposes.  

The main contribution of the present paper is a description of quantum measurement entirely in terms of tensor products, with no recourse to additive structure. The conceptual substance of this description is expressed in the framework of $\dagger$-compact categories through a simple, operationally motivated definition of \em classical objects\em, introduced in our work in 2005, and first presented in print here. A classical  object, as a \em $\dagger$-compact Frobenius algebra\em, equipped with copying and deleting operations, also provides an abstract counterpart to \em GHZ-states\/\em \cite{GHZ0}.  We moreover expose an intriguing conceptual and structural connection between the classical capabilities to copy and delete data, as compared to quantum \cite{NoDe,NoCl}, and the mechanism of quantum measurement: the classical interactions emerge as comonoid homomorphisms, i.e. those morphisms that commute with  copying and deleting.  While each classical object canonically induces a non-degenerate quantum measurement, we show that general quantum measurements arise as \em coalgebras \em for the comonads induced by classical objects.  Quite remarkably, this coalgebra structure exactly captures von Neumann's projection postulate \em in a resource sensitive fashion\em.  Furthermore, the probabilistic content of quantum measurements is then captured using the abstract version of completely positive maps, due to Selinger \cite{Selinger2}.  Using these conceptual components, captured in a succinct categorical signature, we provide a purely graphical derivation of teleportation and dense coding. 

As a first application of the introduced classical structure, we spell out a purely multiplicative form of \em projective quantum measurements\em. In subsequent work \cite{Paquette}, Paquette and the first author extend this treatment to POVMs, and prove \em Naimark's theorem\em \/ entirely within our graphical calculus. Extended abstract \cite{CPP} surveys several important directions and results of further work. The fact that quantum theory can be developed without the additive type constructors suggests a new angle on the question of \em parallelism vs.~entanglement\em. In the final sections of the present paper, we show that superposition too can be described entirely in terms of the monoidal structure, in contrast with the usual Hilbert space view, where entanglement is described as a special case of a superposition.  

%There are also clear structural connections with 
%%certain fields of mathematical physics such as 
%TQFT \cite[and references therein]{Kock,Lauda,Street}. Could categorical quantum semantics provide any physical insights about these mathematical structures? 
% [[this seems out of context. why would connections with TQFT suddenly provide physical insights about "these mathematical structures"? the traffic of insights between math and physics is a story in itself, which has not really been prepared here.]] 

\section{Categorical semantics}\label{sec:AQM}%WAS: Categorical semantics and graphical calculus
In this section, we present both the simple categorical algebra of $\dagger$-compact categories, and the corresponding graphical calculus. In a formal sense, they capture exactly the same structure, and the reader is welcome to pick her favorite flavor (and sort of ignore the other one).   

\subsection{$\dagger$-Compact categories} 

In a \em symmetric monoidal category \em \cite{MacLane} the objects form a monoid with the tensor $\otimes$ as multiplication  and an object $\II$ as the multiplicative unit, up to the coherent\footnote{\em Coherence \em here means that all diagrams composed of these natural transformations commute. In particular, there is at most one natural isomorphism between any two functors composed from $\otimes$ and $\rm I$ \cite{MacLane:coherence}. As a consequence, some functors can be transferred along these canonical isomorphisms, which then become identities. Without loss of generality, one can thus assume that $\alpha$, $\lambda$ and $\rho$ are identities, and that the objects form an actual monoid with $\otimes$ as multiplication and $I$ as unit. Such monoidal categories are called {\em strict}. For every monoidal category, there is an equivalent strict one.} natural isomorphisms
%\[\lambda_A : A \simeq {\rm I}\otimes A\quad\sigma_{A,B}:A\otimes B\simeq B\otimes A\quad\alpha_{A,B,C}:A\otimes(B\otimes C)\simeq (A\otimes B)\otimes C\]
\[
\lambda_A : A \simeq {\rm I}\otimes A\qquad 
\rho_A: A \simeq A\otimes{\rm I}\qquad 
\alpha_{A,B,C}:A\otimes(B\otimes C)\simeq (A\otimes B)\otimes C\,.
\]
The fact that a monoidal category is symmetric means that this monoid is commutative, up to the natural transformation 
\[ 
\sigma_{A,B}:A\otimes B\simeq B\otimes A
\]
coherent with the previous ones. We shall assume that  $\alpha$ is strict, i.e. realized by identity, but it will be convenient to carry $\lambda$ and $\rho$ as explicit structure. Physically, we interpret the objects of a symmetric monoidal category as system types, e.g.~qubit, two qubits, classical data, qubit + classical data etc. A morphism  should be viewed as a \em physical operation\em, e.g.~unitary, or a measurement, classical communication etc.  The tensor captures  \em compoundness \em i.e.~conceiving two systems or two operations as one. Morphisms of type $\II\to A$ represent \em states \em conceived through their respective preparations, whereas morphisms of the type $\II\to\II$ capture  \em scalars \em e.g.~\em probabilistic weights \em --- cf.~complex numbers $c\in\mathbb{C}$ are in bijective correspondence with linear maps 
$\mathbb{C}\to\mathbb{C}::1\mapsto c$.
Details of this interpretation are in  \cite{Cats}.

A symmetric monoidal category is {\em compact\/} \cite{Kelly,KellyLaplaza} if each of its objects has a {\em dual}. An object $B$ is dual to $A$ when it is given with a pair of morphisms
$\eta : \II\to B \otimes A$ and $\varepsilon : A\otimes B \to \II$ often called {\em unit\/} and {\em counit}, satisfying
\beq\label{eq:compact}
(\varepsilon\otimes 1_A)\circ (1_A\otimes \eta) = 1_A \quad{\rm and} \quad(1_B\otimes \varepsilon)\circ (\eta\otimes 1_B) = 1_B\,.
\eeq
It follows that any two duals of $A$ must be isomorphic.\footnote{If $\eta,\varepsilon$ make $B$ dual to $A$, while $\widetilde{\eta},\widetilde{\varepsilon}$ make $\widetilde B$ dual to $A$, then $(1_{\widetilde{B}}\otimes \varepsilon)\circ (\widetilde{\eta}\otimes 1_B):B\to \widetilde B$ and $(1_B\otimes \widetilde\varepsilon)\circ (\eta\otimes 1_{\widetilde{B}}):{\widetilde{B}}\to B$ make $B$ and $\widetilde B$ isomorphic.} A representative of the isomorphism class of the duals of $A$ is usually denoted by $A^\ast$. The corresponding unit and counit are then denoted $\eta_A$ and $\varepsilon_A$.

A {\em symmetric monoidal $\dagger$-category} $\CC$ comes with a  contravariant functor $(-)^\dagger : \CC^{op}\to \CC$, which is identity on the objects, involutive on the morphisms, and preserves the tensor structure \cite{Selinger2}. The image $f^\dagger$ of a morphism $f$ is called its (abstract) {\em adjoint}.   

Finally, $\dagger$-compact categories \cite{AC1,AC1.5} sum up all of the above structure, subject to the additional coherence requirements that
\begin{itemize}
\item every natural isomorphism $\chi$, derived from the symmetric monoidal structure, must be {\em unitary}, i.e. satisfies $\chi^\dagger\circ \chi = 1$ and $\chi\circ \chi^\dagger = 1$, and
\item $\eta_{A^*} = \varepsilon^\dagger_A = \sigma_{A^*A}\circ \eta_A$.
\end{itemize}
Since in a $\dagger$-compact category $\varepsilon_A = \eta_{A^*}^\dagger$ some of the structure of the duals becomes redundant. In particular, it is sufficient to stipulate the units $\eta: \II\to A^\ast\otimes A$, which we call {\em Bell states}, in reference to their physical meaning. In fact, one can skip the above stepwise introduction, and define $\dagger$-compact categories \cite{AC1.5,deLL} simply as a symmetric monoidal category with
\bit
\item an involution
$A\mapsto A^*$,
\item a contravariant, identity-on-objects, monoidal involution
$f\mapsto f^\dagger$, 
\item for each object $A$ a distinguished morphism
$\eta_A:\II\to A^*\otimes A$,
\eit 
which make the diagram
\beq\label{SCCC}
\begin{diagram}A&\lTo^{\simeq}&{\rm I}\otimes
A&\lTo^{\,\eta_{A^*}^\dagger\!\otimes 1_A}&A\otimes 
A^*\otimes A\\
\uTo^{1_A}&&&&\uTo_{1_{A\otimes A^*\otimes A}}\\ 
A&\rTo_{\simeq}&A\otimes{\rm
I}&\rTo_{1_A\otimes\eta_A}&A\otimes A^*\otimes A
\end{diagram}
\eeq
commute. In a sense, $\dagger$-compact categories can thus be construed as an abstract axiomatization of the Bell-states, familiar in the Hilbert space formalism
\[
\eta_{\cal H}:\mathbb{C}\to {\cal H}^*\otimes{\cal H}::1\mapsto \sum_{i\in I}|\, i\,i\,\rangle\,,
\]
where ${\cal H}^*$ is the \em conjugate space \em to ${\cal H}$. This apparently simple axiomatics turns out to generate an amazing amount of the Hilbert space machinery, including the Hilbert-Schmidt inner-product, completely positive map and POVMs \cite{AC1.5,deLL,Paquette,Selinger2}, to mention just a few.  
  
\subsection{Graphical calculus} 

The algebraic structure of $\dagger$-compact categories satisfies exactly those equations that can be proven its \em graphic language\em, which we shall now describe. In other words, the morphisms of the free $\dagger$-compact category can be presented as the well-formed diagrams of this graphic language. Proving such statements, and extracting sound and complete graphic languages for particular categorical varieties, has a long tradition in categorical algebra \cite{MacLane:coherence,Kelly,KellyLaplaza,JS,JSV}. Using the deep results about coherent categories in particular \cite{KellyLaplaza}, Selinger has elegantly derived a succinct coherence argument for the graphic language of $\dagger$-compact categories in \cite[{\bf Thm}.~3.9]{Selinger2}. We briefly summarize a version of this graphic language. 

The objects of a $\dagger$-compact category are represented by tuples of wires, whereas the morphisms are the I/O-boxes. Sequential composition connects the output wires of one box with the input wires of the other one. The tensor product is the union of the wires, and it places the boxes next to each other.  A physicist-friendly introduction to this graphical language for symmetric monoidal categories is in \cite{Cats}.  The main power of the graphical language lies in its representation of duality. The Bell state (unit) and its adjoint (counit) correspond to a wire from $A$ returning into $A^\ast$, with the directions reversed:
\par\noindent
\begin{minipage}[b]{1\linewidth}  
\centering{\epsfig{figure=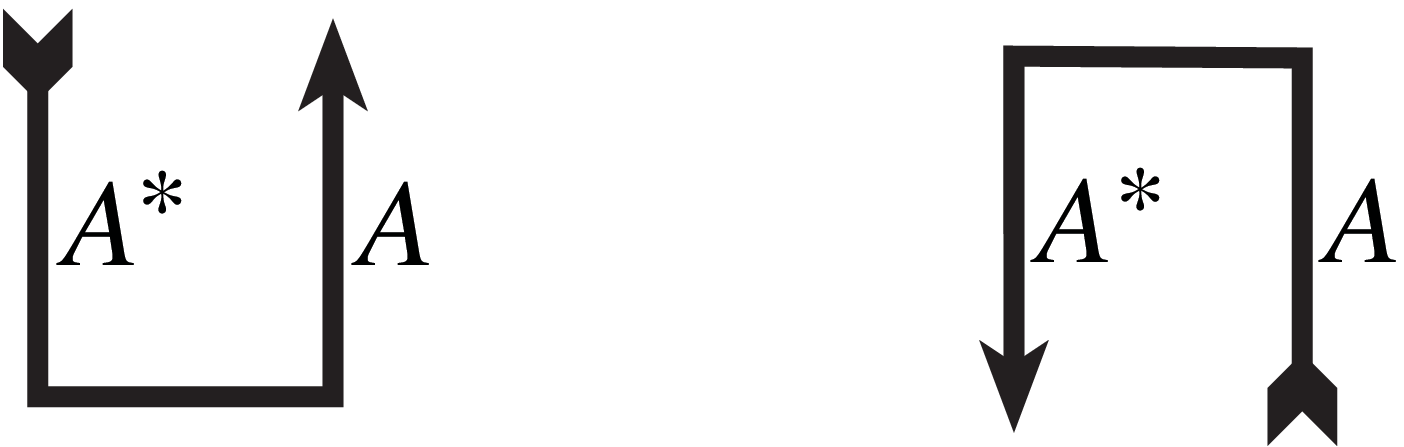,width=124pt}\ }  
\end{minipage}

%The commutative diagram that establishes that $A^\ast$ is dual to $A$ can be drawn in the form that we shall see repeated several times:
Graphically, the composition of $\eta_A$ and $\varepsilon_A = \eta^\dagger_{A^*}$ as expressed in commutative diagram (\ref{SCCC}) boils down to
\par\noindent
\begin{minipage}[b]{1\linewidth}  
\centering{\epsfig{figure=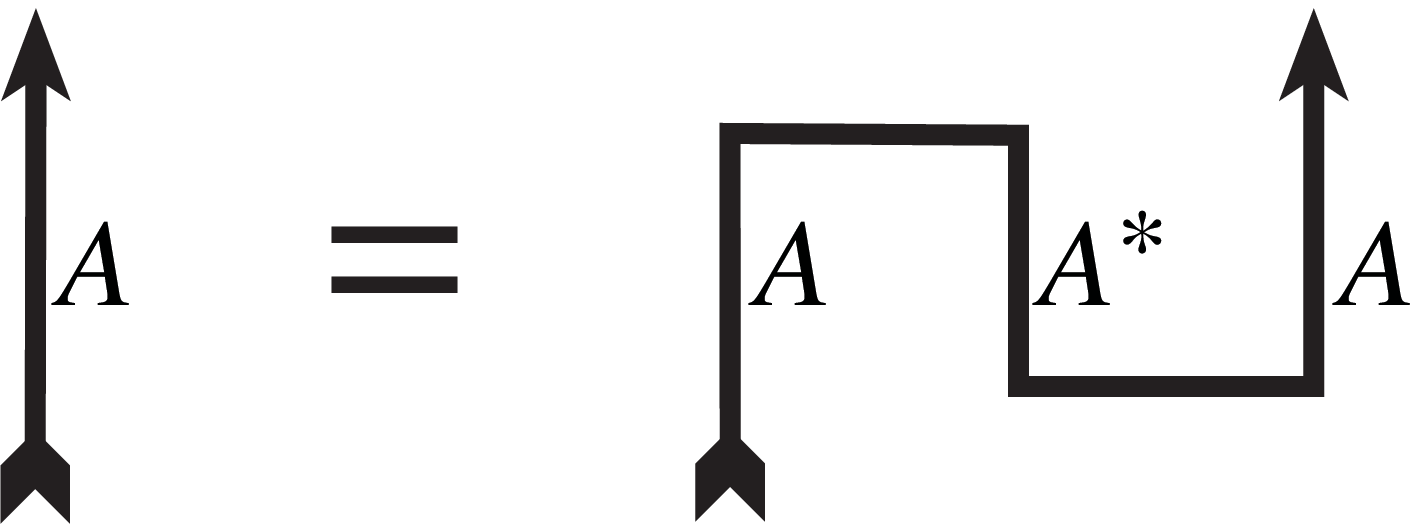,width=128pt}\ }  
\end{minipage}
%\par\vspace{2mm}\par 
Note that in related papers such as \cite{Kindergarten} a more involved notation
\par\noindent
\begin{minipage}[b]{1\linewidth}  
\centering{\epsfig{figure=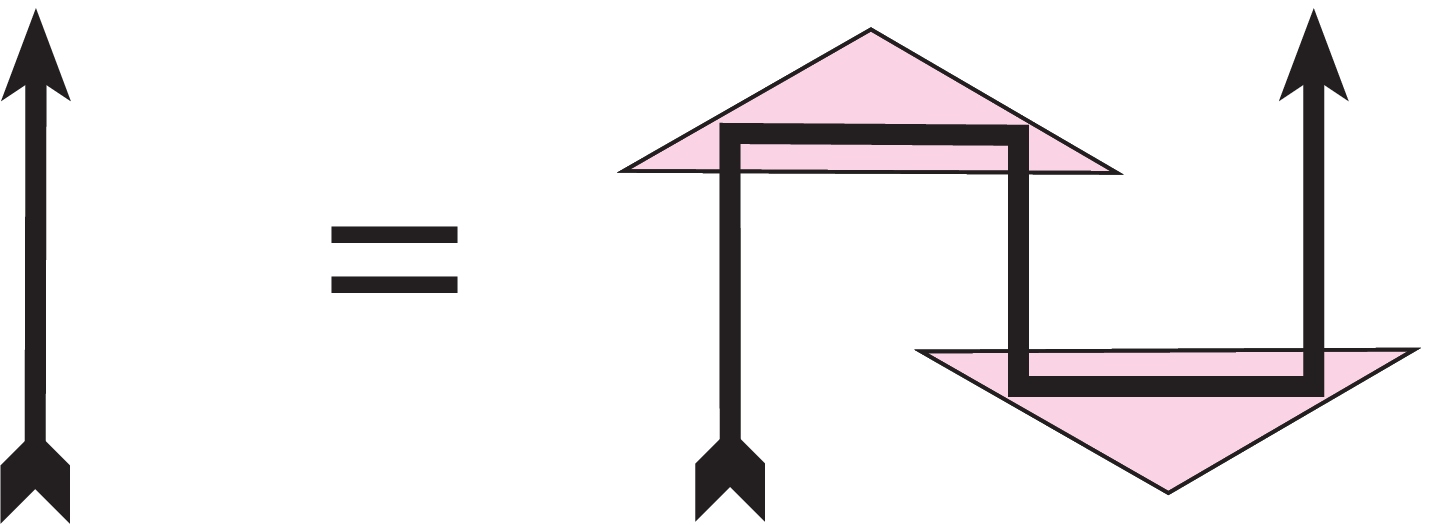,width=128pt}\ }  
\end{minipage}
appears. The triangles witness the fact that in physical terms
\par\noindent
\begin{minipage}[b]{1\linewidth}  
\centering{\epsfig{figure=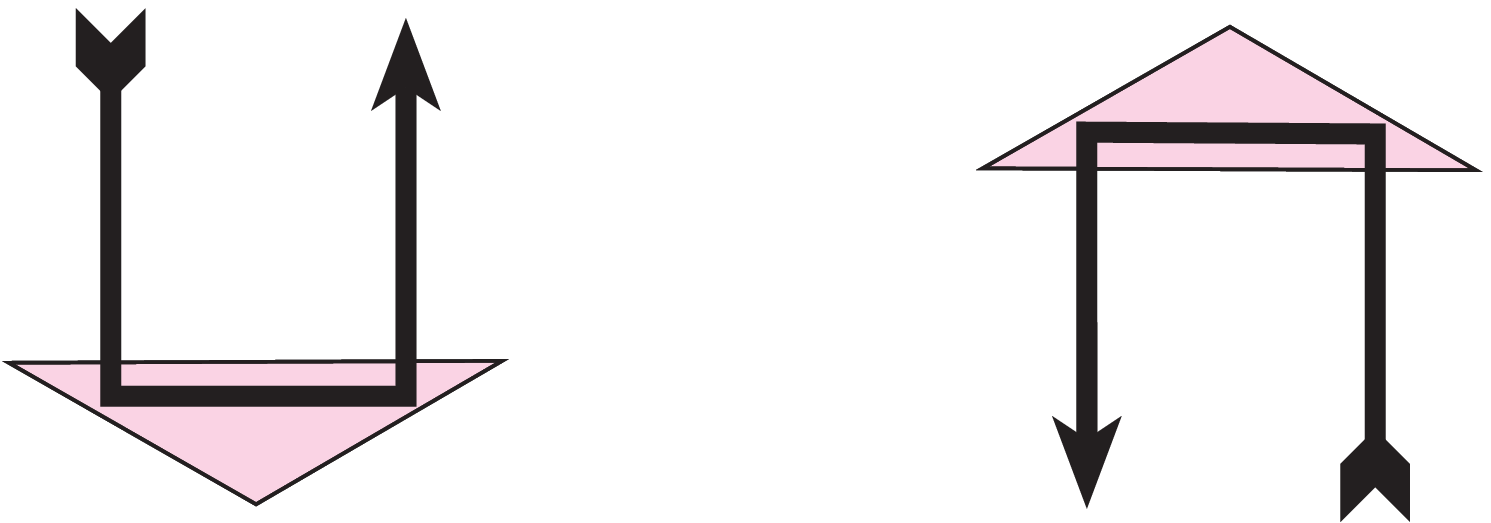,width=144pt}\ }  
\end{minipage}
respectively stand for a preparations procedure, or state, or \em ket\em, and for the corresponding \em bra\em, with an inner-product or \em bra-ket \em 
\par\noindent
\begin{minipage}[b]{1\linewidth}  
\centering{\epsfig{figure=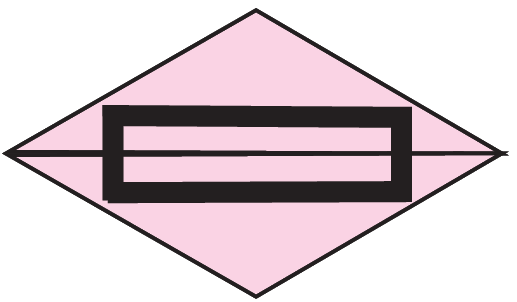,width=60pt}\ \ \ }  
\end{minipage}
then yielding a \em diamond shaped scalar \em (cf.~\cite{Kindergarten}), while the wire itself is now a \em loop\em.  In this paper we will omit these special bipartite triangles.

Given a choice of the duals $A \mapsto A^\ast$, one can follow the same pattern to define the arrow part $f\mapsto f^\ast$ of the duality functor $(-)^\ast : \CC^{op}\to \CC$ by the commutativity of the following diagram:   
\begin{diagram}A^*&\lTo^{\simeq}&A^*\otimes {\rm I}&\lTo^{1_{A^*}\otimes
\eta^\dagger_{B^*}}&A^*\otimes
B\otimes B^*\\
\uTo^{f^*}&&&&\uTo_{1_{A^*}\!\otimes f\otimes 1_{B^*}\hspace{-1.3cm}}\\
B^*&\rTo_{\simeq}&{\rm I}\otimes B^*&\rTo_{\eta_A\otimes
1_{B^*}}&A^*\otimes 
A\otimes B^*
\end{diagram}
Replacing $f:A\to B$ by $f^\dagger : B\to A$, we can similarly define $f_\ast : A^\ast \to B^\ast$, and thus extend the duality assignment $A\mapsto A^\ast$ by the morphism assignment $f\mapsto f_\ast$ to the covariant functor $(-)_\ast : \CC \to \CC$. It can be shown \cite{AC1} that the adjoint decomposes in every $\dagger$-compact category as
\[
f^\dagger=(f^*)_*=(f_*)^*
\]
with both $(-)^*$ and $(-)_*$ involutive. In finite dimensional Hilbert spaces and linear maps {\bf FdHilb}, these two functors respectively correspond to \em transposition \em and \em complex conjugation\em. The functor $(-)_\ast : \CC \to \CC$ will thus be called {\em conjugation}; the image $f_*$ is a conjugate of $f$. Graphically, the above diagram defining $f^\ast$, and the similar one for $f_\ast$, respectively become
\par\vspace{2mm}\par\noindent  
\begin{minipage}[b]{1\linewidth}
\centering{\epsfig{figure=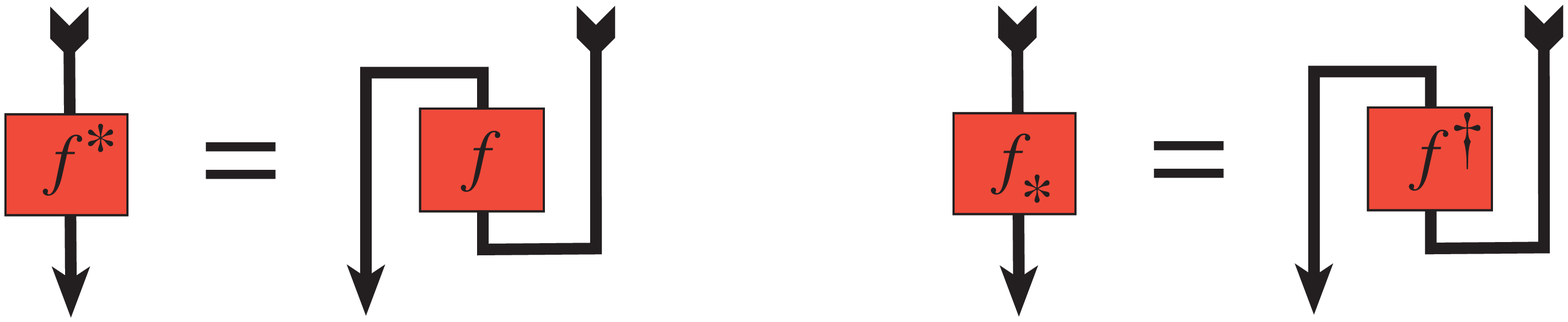,width=260pt}}          
\end{minipage}
\par\vspace{2mm}\par\noindent
The direction of the arrows is, of course, just relative, and we have chosen to direct the arrows down in order to indicate that both $f^\ast$ and  $f_\ast$ have the duals as their domain and codomain types. We will use horizontal reflection to depict $(-)^\dagger$ and Selinger's 180$^\circ$ rotation \cite{Selinger2} to depict  $(-)_*$,  resulting in:
\par\vspace{2mm}\par\noindent
\begin{minipage}[b]{1\linewidth}
\centering{\epsfig{figure=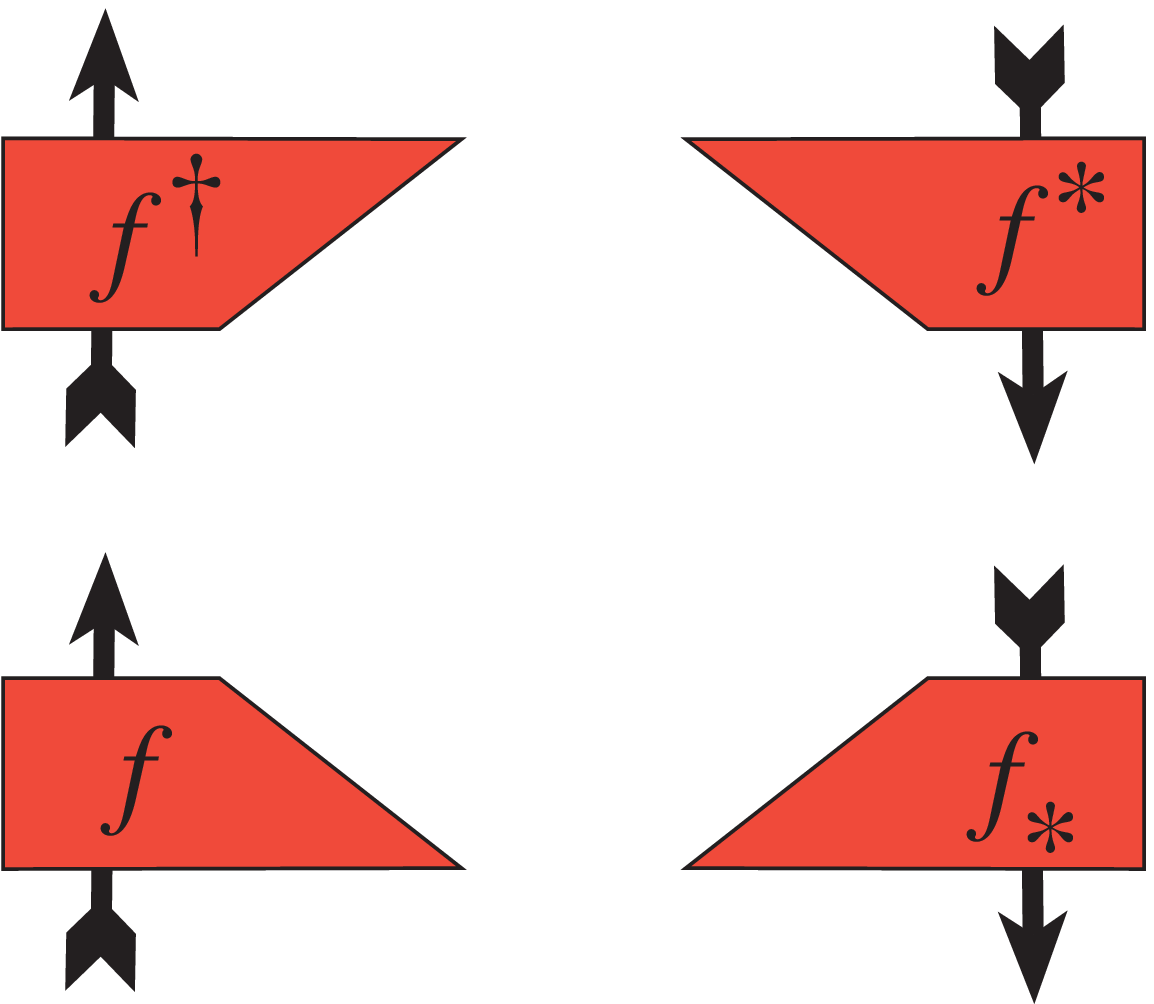,width=110pt}}        
\end{minipage}

\subsection{Scalars, trace, and partial transpose}

One can prove that the monoid  $\CC(\II,\II)$ is always commutative \cite{KellyLaplaza} and induces  a \em scalar
multiplication \em
\[
s\bullet f:=\lambda_B^{-1}\circ (s\otimes f)\circ\lambda_A:A\to B
\]
which by naturality satisfies 
\beq\label{eq:scalmult}
(s\bullet f)\circ (t\bullet g)=(s\circ t)\bullet(f\circ g)
\quad
(s\bullet f)\otimes (t\bullet g)=(s\circ  
t)\bullet(f\otimes g)\,.
\eeq
As already indicated above, we will depict these scalars by diamonds, and such scalars can arise as loops. The equations (\ref{eq:scalmult}) show that these diamonds capturing probabilistic weights can be `freely moved in the pictures'. 

The compact structure of $\dagger$-compact categories induces the familiar \em trace \em operation \cite{JSV}, 
\[
{\rm tr}^{C}_{A,B}:{\bf C}(C\otimes A,C\otimes B)\to{\bf C}(A,B)
\]
which maps $f:C\otimes A\to C\otimes B$ to
\begin{diagram}B&\lTo^{\!\!\!\!\!\!\simeq\!\!}&{\rm I}\otimes
B&\lTo^{\eta_C^\dagger\otimes 1_B}&C^*\!\otimes C\otimes B\\
\uTo^{{\rm tr}^C_{A,B}(f)}&&&&\uTo_{1_{C^*}\!\otimes f}\\
A&\rTo_{\!\!\!\!\!\!\simeq\!\!}&{\rm I}\otimes
A&\rTo_{\eta_C\otimes 1_A}& C^*\!\otimes C \otimes A
\end{diagram}
The graphic form of ${\rm tr}^C_{A,B}(f)$ is:
\par\vspace{2mm}\par\noindent
\begin{minipage}[b]{1\linewidth}
\centering{\epsfig{figure=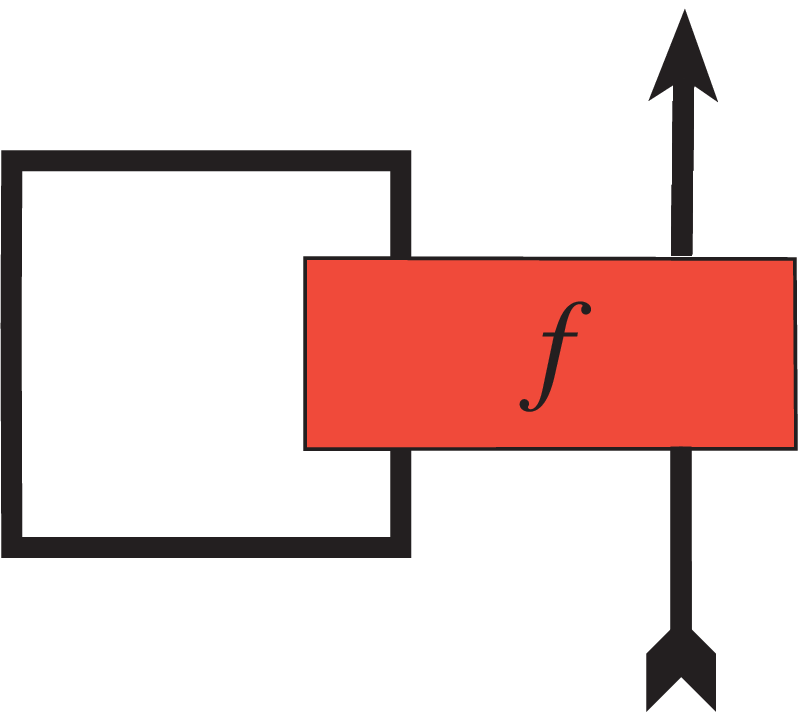,width=75pt}}     
\end{minipage}
A less familiar operation is \em partial transpose \em
\[
{\rm pt}^{C,D}_{A,B}:{\bf C}(C\otimes A,D\otimes B)\to{\bf C}(D^*\otimes A,C^*\otimes B)
\]
which maps $f:C\otimes A\to D\otimes B$
\begin{diagram}C^*\!\otimes B&\lTo^{\!\!\!\!\!\!\simeq\!\!}&C^*\!\otimes{\rm I}\otimes
B&\lTo^{1_{C^*}\!\otimes\eta_D^\dagger\otimes 1_B\!\!}&C^*\!\otimes D^*\otimes D\otimes B\\
\uTo^{{\rm pt}^{C,D}_{A,B}(f)}&&&&\uTo_{\sigma_{D^*\!,C^*}\!\otimes f}\\
D^*\!\otimes A&\rTo_{\!\!\!\!\!\!\simeq\!\!}&D^*\!\otimes {\rm I}\otimes
A&\rTo_{1_{D^*}\!\otimes\eta_C\otimes 1_A\!\!}&D^*\!\otimes C^*\!\otimes C\otimes A
\end{diagram}
In a picture, ${\rm pt}^{C,D}_{A,B}(f)$ is:
\par\vspace{2mm}\par\noindent
\begin{minipage}[b]{1\linewidth}
\centering{\epsfig{figure=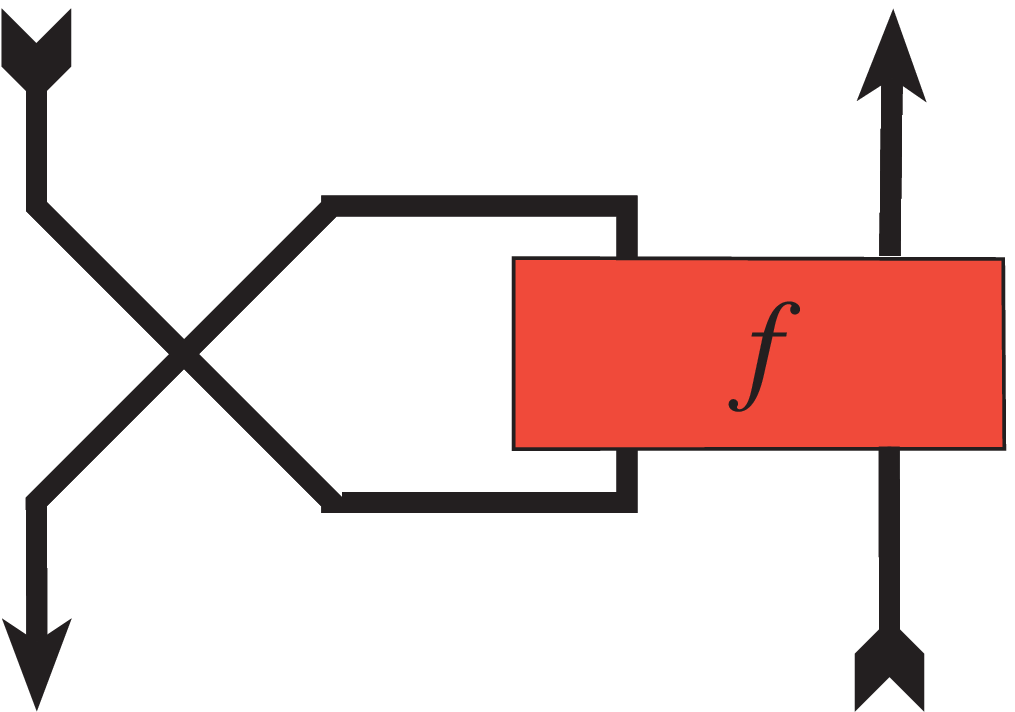,width=90pt}}        
\end{minipage}
\par\vspace{2mm}\par\noindent
Partial transpose can in fact be viewed as an internalisation of the swap-actions $\sigma_{C,D}\circ-$ and $-\circ\sigma_{C,D}$, combined with a transposition of the dual space, so that it does not swap two inputs, or two outputs, but an input and an output.

\section{Sums and bases in Hilbert spaces}\label{sec:crux}%WAS: Remarks about sums, copying and bases in Hilbert spaces

To motivate the algebraic and diagrammatic analysis of quantum measurement in the next section, we first discuss some particular aspects of the Hilbert space model of quantum mechanics. 

\subsection{Sums in quantum mechanics}  

Sums occur in the Hilbert space formalism both as a part of the linear structure of states, as well as a part of their projective (convex) structure, through the fundamental theorem of projective geometry and Gleason's theorem \cite{Piron}. Viewed categorically, these structures lift, respectively, to a vector space enrichment and a projective space enrichment of operators, typically yielding a $C^*$-algebra.  They appear to be necessary because of the specific nature of quantum measurement, and the resulting quantum probabilistic structure.  The additive structure permeates not only states, but also state spaces; it is crucial not only for adding vectors, but also for composing and decomposing spaces.  
In fact, one verifies that operator sums arise from the direct sum:
\begin{diagram}
\mathbb{C}^{\oplus n}&\rTo{f+g}&\mathbb{C}^{\oplus m}\\
\dTo^d&&\uTo_{d^\dagger}\\
\mathbb{C}^{\oplus n}\oplus\mathbb{C}^{\oplus n}&
\rTo_{\ \ f\oplus g\ \ }&\mathbb{C}^{\oplus m}\oplus \mathbb{C}^{\oplus m}
\end{diagram}
where $d::|\,i\rangle\mapsto |\,i\rangle\oplus|\,i\rangle$ is the additive diagonal.  As a particular case we have that the vector sums arise from 
\begin{diagram}
\mathbb{C}&\rTo{|\psi\rangle+|\phi\rangle}&\mathbb{C}^{\oplus n}\\
\dTo^d&&\uTo_{d^\dagger}\\
\mathbb{C}\oplus\mathbb{C}&
\rTo_{\ \ |\psi\rangle\oplus|\phi\rangle\ \ }&\mathbb{C}^{\oplus n}\oplus \mathbb{C}^{\oplus n}
\end{diagram}
where $|\psi\rangle,|\phi\rangle:\mathbb{C}\to\mathbb{C}^{\oplus n}$, 
recalling that vectors $|\psi\rangle\in \mathbb{C}^{\oplus n}$ are indeed, by linearity,  in bijective correspondence with the linear maps 
\[
\mathbb{C}\to \mathbb{C}^{\oplus n}::1\mapsto |\psi\rangle\,.
\]
In addition to this, the direct sum canonically also defines bases  (cf.~the computational base) in terms of the $n$ canonical injections
\[
\mathbb{C}\hookrightarrow\mathbb{C}^{\oplus n}::1\mapsto(0,\ldots,0,1,0,\ldots,0)\,. 
\]

\subsection{No-Cloning and existence of a natural diagonal} 

The classic No-Cloning Theorem \cite{NoCl} states that there exists no unitary operation  
\beq\label{eq:cloning}
Clone:{\cal H}\otimes{\cal H}\to{\cal H}\otimes{\cal H}::|\psi\rangle\otimes|\,0\rangle\mapsto|\psi\rangle\otimes|\psi\rangle\,.
\eeq
In categorical terms, this means that there is no \em natural\em\/ diagonal for the Hilbert space tensor product. Formally, a diagonal is a family of linear maps
\[
\Delta_{\cal H}:{\cal H}\to{\cal H}\otimes{\cal H}\,,
\]
one for each of Hilbert space ${\cal H}$. Such a family is said to be natural if for every linear map $f:{\cal H}\to{\cal H}'$ the diagram
\beq\label{diag:DiagCom}\begin{diagram}
{\cal H}&\rTo^{f}&{\cal H}'\\
\dTo^{\Delta_{\cal H}}&&\dTo_{\Delta_{{\cal H}'}}\\
{\cal H}\otimes {\cal H}&\rTo_{f\otimes f}&{\cal H}'\otimes {\cal H}'\\ 
\end{diagram}\eeq
commutes.  For instance, the family
\[
\delta:{\cal H}\to{\cal H}\otimes{\cal H}::|\,i\,\rangle\mapsto |\,ii\,\rangle
\]
is a diagonal, but the No-Cloning Theorem implies that it cannot be natural. Indeed, it is not hard to see that the above diagram fails to commute, say, for ${\cal H}:=\mathbb{C}$,  ${\cal H}':=\mathbb{C}\oplus\mathbb{C}$ and $f:1\mapsto |\,0\rangle+|\,1\rangle$.  
%
%We will now show that existence of a `cloning machine' as in (\ref{eq:cloning}) implies existence of a natural diagonal. Hence absence of a  natural diagonal has the No-Cloning theorem as a corollary. Moreover, unitarity of the cloning machine does not play a role in this argument, so in fact we obtain a stronger No-Cloning theorem.  Given a (not necessarily unitary) map as in (\ref{eq:cloning}) consider 
In general, given a cloning machine (\ref{eq:cloning}), one can define  a natural diagonal
\[
\Delta_{\cal H}:=Clone\circ(-\otimes|\,0\rangle):{\cal H}\to{\cal H}\otimes{\cal H}::
|\psi\rangle\mapsto|\psi\rangle\otimes|\psi\rangle\,.
\]
To prove its naturality, note that (\ref{eq:cloning}) holds for every  $|\psi\rangle$, including $|\psi\rangle =|f(\varphi)\rangle$, %with $f$ arbitrary,  
which gives
\beqa
(\Delta_{\cal H}\circ f)(|\varphi\rangle)&\!\!=\!\!&
\Delta_{\cal H}(|f(\varphi)\rangle)=|f(\varphi)\rangle\otimes|f(\varphi)\rangle
=(f\otimes f)(|\varphi\rangle\otimes|\varphi\rangle)\\
&\!\!=\!\!&(f\otimes f)(\Delta_{\cal H}(|\varphi\rangle))=((f\otimes f)\circ\Delta_{\cal H})(|\varphi\rangle)
\eeqa 
which shows that diagram (\ref{diag:DiagCom}) commutes. 

A diligent reader may have noticed that commutativity of  (\ref{diag:DiagCom}) actually implies that a diagonal $\Delta$ must be independent on the bases, because a change of base can be viewed as just another linear map $f$ (e.g.~\cite{Cats}).  In fact, invariance under the base change was one of the original motivations behind the categorical concept of naturality, viz. of natural transformations \cite{MacLane}.

\subsection{Measurement and bases} 

When diagonalized, self-adjoint operators, which represent measurements in quantum mechanics
%, if non-degenerated,\footnote{We only make this assumption for the sake of simplicity of the argument.} 
boil down, modulo a change of base, to two families of data:   eigenvalues and eigenvectors.  Viewed quantum informatically, eigenvalues are merely token witnesses which discriminate outcomes. A non-degenerate measurement thus essentially corresponds to a base, and a degenerate one can also be captured by a base, 
%provided that we also provide 
and an equivalence relation over it.  Taking another look at the map $|\,i\,\rangle\stackrel{\delta}{\mapsto} |\,ii\,\rangle$.\footnote{This map, when assigning agents i.e.~$|\,i\,\rangle_A\stackrel{\delta}{\mapsto} |\,i\,\rangle_A\otimes |\,i\,\rangle_B$,  has appeared in the literature  under the name \em coherent bit\em, as a `between classical and quantum'-channel \cite{Devetak,Harrow}.  A more detailed study of this connection can be found in \cite{CPP}.}, we see that it \em does copy \em the base vectors, but {\em not\/} other states:  
\[
|\psi\rangle=\sum_ic_i|\,i\,\rangle\stackrel{\delta}{\ \mapsto\ }\sum_ic_i  |\,ii\,\rangle\not=|\psi\rangle\otimes|\psi\rangle\,.
\]
This map, in fact, exactly captures the base $\{|\,i\,\rangle\}_i$, because
\[
\delta::\sum_{i\in I}c_i|\,i\,\rangle\ \mapsto\ \sum_{i\in I}c_i |\,ii\,\rangle\,,
\]
yields a disentangled state if and only if the index set $I$ is a singleton, i.e.~if and only if the linear combination $\sum_{i\in I}c_i|\,i\,\rangle$ boils down to a base vector.  Going in the opposite direction, we can also recover the base as the image of pure tensors under the map\footnote{This operation $\delta^\dagger:{\cal H}\otimes{\cal H}\to{\cal H}$ has appeared in the quantum informatics literature under the name \em fusion\em, providing a means for constructing cluster states \cite{fusion2,fusion1}.}
\[
\delta^\dagger::
\left\{
\begin{array}{lcccc}
i\not=j&:& |\,ij\,\rangle&\mapsto& \vec{o}\vspace{1mm}\\
else&:&  |\,ii\,\rangle&\mapsto & |\,i\,\rangle
\end{array}
\right.
\]
Since the linear diagonal $\delta$ thus captures the base, it is of course not independent of the base, and cannot be a natural transformation, in the categorical sense. We shall see that its importance essentially arises from this "un-naturality". Restricted to the base vectors, $\delta$ is a `classical' \em copying \em operation par excellence; viewed as a linear operation on all of the Hilbert space, it drastically fails naturality tests. 

The upshot is that, this operation allows us to characterize \em classical measurement context \em as the domain where it faithfully copies data, with no recourse to an explicit base. If needed, however, the base can be extracted from among the quantum states as consisting of just those vectors that can be copied.

\subsection{Vanishing of non-diagonal elements and deletion} 

The map $\delta$ also allows capturing the `formal  decohering' in quantum measurement, i.e.~the vanishing of the non-diagonal elements in the passage of the initial state represented as a density matrix within the measurement base to the density matrix describing  the resulting ensemble of possible outcome states.\footnote{See \cite{GisinPiron} for a discussion why we call this `formal decohering'.} Indeed, non-diagonal elements get erased setting
\[
\delta\circ\delta^\dagger::\left\{
\begin{array}{lcccccc}
i\not=j&:& |\,ij\,\rangle&\mapsto& \vec{o}&\mapsto& \vec{o}\vspace{1mm}\\
else&:&  |\,ii\,\rangle&\mapsto & |\,i\,\rangle&\mapsto  &|\,ii\,\rangle
\end{array}
\right.
\]

Note also that $\delta$'s adjoint $\delta^\dagger$ doesn't delete classical data, but \em compares \em its two inputs and only passes on data if they coincide.  \em Deletion \em is 
\[
\epsilon::|\,i\,\rangle\mapsto 1\qquad{\rm that\ is}\qquad
1\otimes \epsilon::
|\,ij\,\rangle\mapsto |\,i\,\rangle\,.
\]
What $\epsilon$ and $\delta^\dagger$ do have in common is the fact that 
\[
\delta^\dagger\circ\delta= (1\otimes \epsilon)\circ\delta:: |\,i\,\rangle\mapsto  |\,ii\,\rangle\mapsto |\,i\,\rangle\,.  
\]

Also, since in Dirac-notation we have $\delta=\sum_i|\,ii\rangle\langle i\,|$, the (base-depen\-dent)
isomorphism $\theta::|\,i\rangle\mapsto \langle i\,|$ applied to the \em bra \em turns $\delta$ into the generalized {\sf GHZ}-state $\sum_i|\,iii\rangle$\/ \cite{GHZ} exposing that $\delta$ is `up to $\theta$' symmetric in all variables.

\subsection{Canonical bases} 

While all Hilbert spaces of the same dimension are obviously isomorphic, they are not all equivalent. Indeed, above we already mentioned that the direct sum structure provides the Hilbert space $\mathbb{C}^{\oplus n}$ with a canonical base, from which it also follows that it is canonically isomorphic to its conjugate space $\bigl(\mathbb{C}^{\oplus n}\bigr)^\ast=\left(\mathbb{C}^\ast\right)^{\oplus n}$, namely for the isomorphism 
\[
\mathbb{C}^{\oplus n}\to \left(\mathbb{C}^\ast\right)^{\oplus n}::
(c_1,\ldots,c_n)\mapsto(\bar{c}_1,\ldots,\bar{c}_n)\,.
\]
In fact, one should not think of $\mathbb{C}^{\oplus n}$ as just being a Hilbert space, but as the pair consisting of a Hilbert space ${\cal H}$ and a base $\{|\,i\,\rangle\}_{i=1}^{i=n}$, which by the above discussion boils down to the pair consisting of a Hilbert space ${\cal H}$ and a linear map $\delta:{\cal H}\to{\cal H}\otimes{\cal H}$ satisfying certain properties, in particular, its matrix being self-transposed in the canonical base. Below, we will assume the correspondence between $\mathbb{C}^{\oplus n}$ and its dual to be strict, something which can always be established by standard methods.  The special status of the objects $\mathbb{C}^{\oplus n}$ in {\bf FdHilb}, in category-theoretic terms, is due to the fact 
the direct sum is both a product and a coproduct and $\mathbb{C}$ the tensor unit \cite{AC1}.

%In fact something more needs to be added to the quantum formalism to be able to describe protocols in a contemporary CS-fashion, namely a classical control structure.  This is mainly the subject of some current \em quantum programming language \em literature, with languages for example proposed by Selinger and Alterkirch \& Grattage, surveyed in \cite{Gay}. In \cite{AC1} a minimal but sufficient classical control structure was provided by distributivity of the additive biproduct over the multiplicative tensor.

\section{Classical objects}\label{sec:CO}

Consider a quantum measurement.  It takes a quantum state as its input and produces a measurement outcome together with a quantum state, which is typically different from the input state due to the \em collapse\em.  Hence the type of a quantum measurement should be
\[
{\cal M}:A\to X\otimes A
\]
where $A$ is of the type \em quantum state \em while $X$ is of the type \em classical data\em.  
But how do we distinguish between classical and quantum data types?  

We will take a very operational view on this matter, and define classical data types as objects which come together with a copying operation
\[
\delta_{(X)}:X\to X\otimes X
\]
and a deleting operation  
\[
\epsilon_{(X)}:X\to\II\,,
\]
counterfactually exploiting the fact that such operations do not exist for quantum data. We will refer to these structured objects $(X,\delta,\epsilon)$ as \em classical objects\em.  The axioms which we require the morphisms $\delta$ and $\epsilon$ to satisfy are motivated by the operational interpretation of $\delta$ and $\epsilon$ as copying and deleting operations of classical data.  This leads us to introducing the notion of a \em special $\dagger$-compact Frobenius algebra\em, which refines the usual topological quantum field theoretic notion of a normalized special Frobenius algebra \cite{Kock}. The defining equality is due to Carboni and Walters \cite{CarboniWalters}.\footnote{They introduced it as a characteristic categorical property of relations.  The connection between the work presented in this paper and Carboni and Walters' categories of relations is in \cite{CPP}.}

\subsection{Special $\dagger$-compact Frobenius algebras}  

An \em internal monoid $(X,\mu,\nu)$ \em in a monoidal category $({\bf C},\otimes, \II)$ is a pair of morphisms
\begin{diagram}
\hspace{-8mm}X\otimes X&\rTo^\mu&X&\lTo^\nu&\II\,,
\end{diagram}
called the \em multiplication \em and the \em multiplicative unit\em,
such that
\begin{diagram}
X&\lTo^{\mu}&X\otimes X&&&&X&&\\
\uTo^{\mu}&&\uTo_{1_X\otimes \mu}&&&\ruTo^{\lambda_X^{-1}}&\uTo~\mu&\luTo^{\rho_X^{-1}}&\\
X\otimes X&\lTo_{\mu\otimes 1_X}&X\otimes X\otimes X&&\II\otimes X&\rTo_{\nu\otimes 1_X}&X\otimes X&\lTo_{1_X\otimes \nu}&X\otimes\II
\end{diagram}
commute. Dually,  an \em internal comonoid $(X,\delta,\epsilon)$ \em  is a pair of morphisms
\begin{diagram}
\hspace{-8mm}X\otimes X&\lTo^\delta&X&\rTo^\epsilon&\II\,,
\end{diagram}
the \em comultiplication \em and the \em comultiplicative unit\em,
such that
\begin{diagram}
X&\rTo^{\delta}&X\otimes X&&&&X&&\\
\dTo^{\delta}&&\dTo_{1_X\otimes \delta}&&&\ldTo^{\lambda_X}&\dTo~\delta&\rdTo^{\rho_X}&\\
X\otimes X&\rTo_{\delta\otimes 1_X}&X\otimes X\otimes X&&\II\otimes X&\lTo_{\epsilon\otimes 1_X}&X\otimes X&\rTo_{1_X\otimes \epsilon}&X\otimes\II
\end{diagram}
commute. Graphically these conditions are:
\par\vspace{2mm}\par\noindent
\begin{minipage}[b]{1\linewidth}
\centering{\epsfig{figure=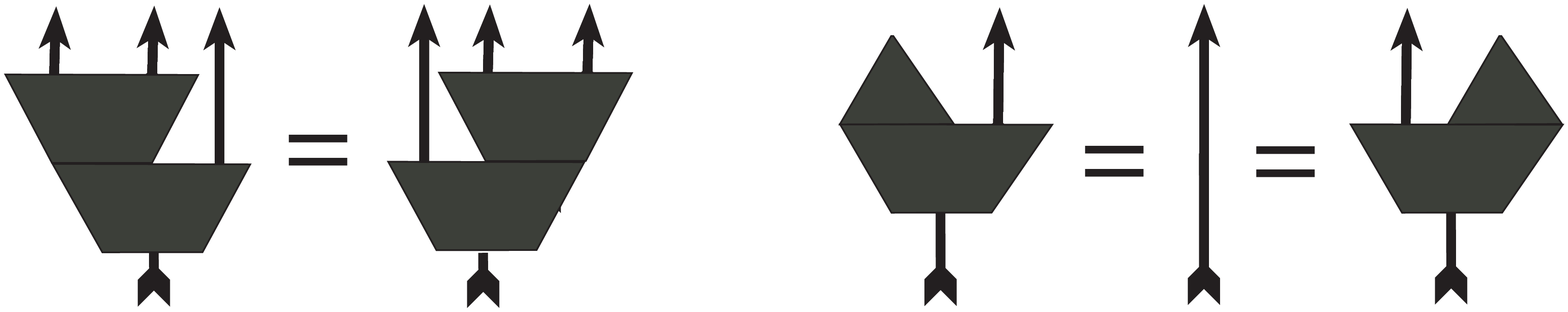,width=270pt}}     
\end{minipage}
\par\vspace{2mm}\par\noindent 
When $({\bf C},\otimes,\II)$ is \em symmetric\em, the monoid is commutative iff $\mu\circ\sigma_{X,X}=\mu$, and the comonoid is commutative iff $\sigma_{X,X}\circ \delta=\delta$, in a picture:
\par\vspace{2mm}\par\noindent
\begin{minipage}[b]{1\linewidth}
\centering{\epsfig{figure=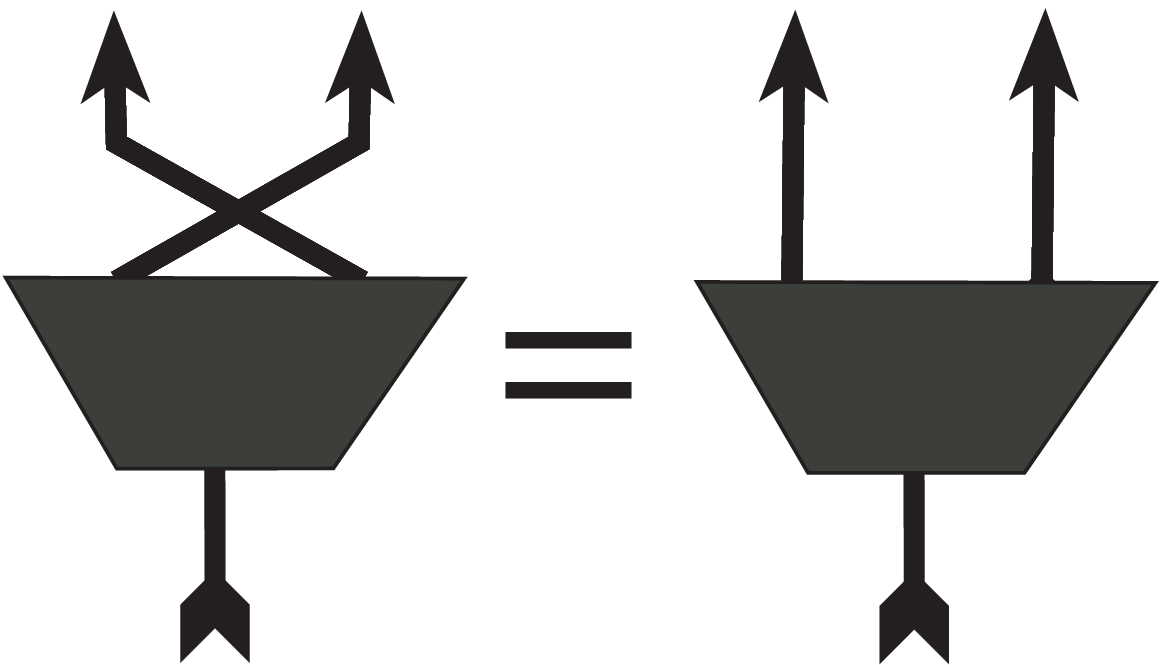,width=96pt}}     
\end{minipage}
\par\vspace{2mm}\par\noindent 
Note that the conditions defining an internal commutative comonoid are indeed what we expect a copying and deleting operation to satisfy.

A \em symmetric Frobenius algebra \em is an internal  commutative monoid $(X,\mu,\nu)$ together with an internal commutative comonoid $(X,\delta,\epsilon)$ which satisfy
\beq\label{eq:Frobenius}
\delta\circ\mu=(\mu\otimes 1_X)\circ(1_X\otimes \delta)\,, 
\eeq
that is, in a picture:
\par\vspace{2mm}\par\noindent
\begin{minipage}[b]{1\linewidth}
\centering{\epsfig{figure=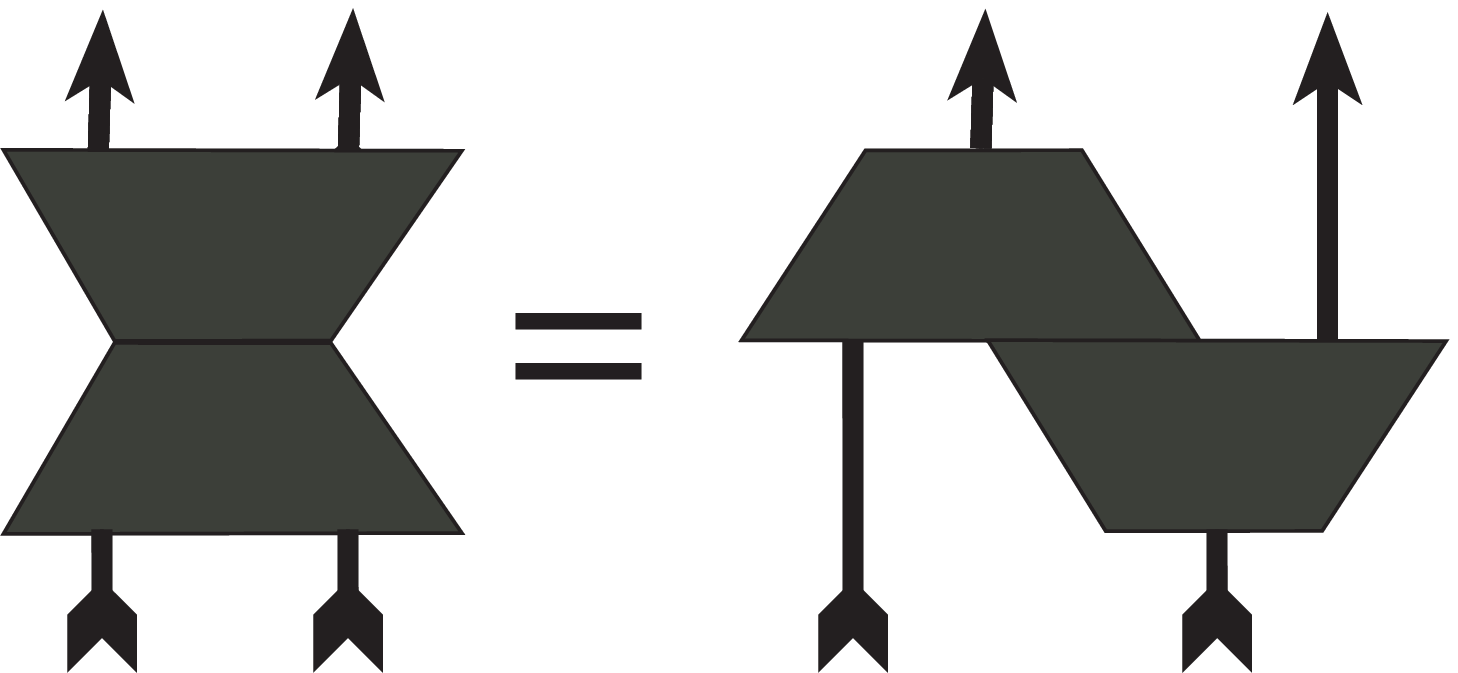,width=108pt}}        
\end{minipage}
\par\vspace{2mm}\par\noindent
It is moreover \em special \em iff $\mu\circ\delta=1_X$, in a picture:
\par\vspace{2mm}\par\noindent
\begin{minipage}[b]{1\linewidth}
\centering{\epsfig{figure=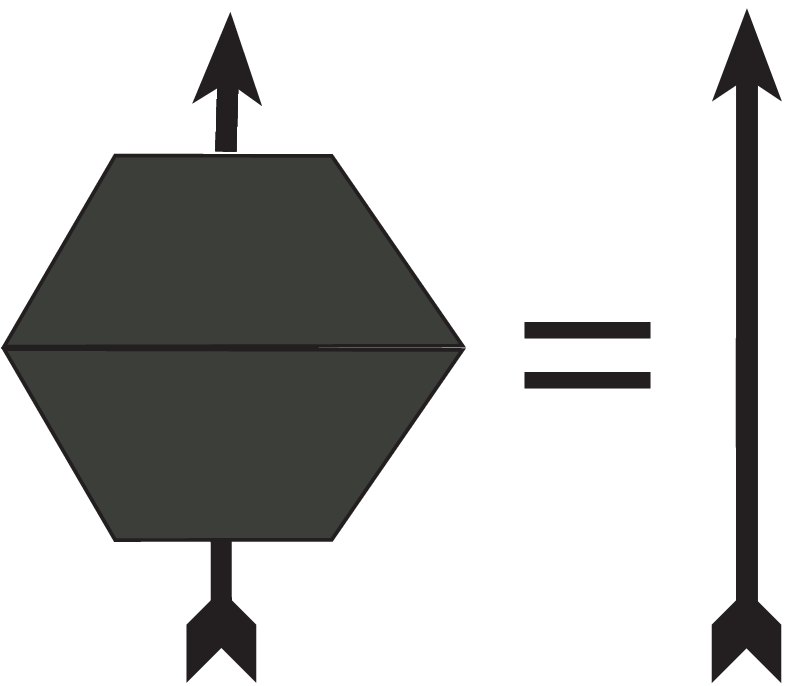,width=63pt}}     
\end{minipage}
\par\vspace{2mm}\par\noindent 
In a symmetric monoidal $\dagger$-category every internal commutative comonoid $(X,\delta,\epsilon)$ also defines an internal commutative monoid $(X,\delta^\dagger,\epsilon^\dagger)$, yielding a notion of \em $\dagger$-Frobenius algebra \em $(X,\delta,\epsilon)$ in the obvious manner.  In such a $\dagger$-Frobenius algebra we have:
\par\vspace{1mm}\par\noindent
\begin{minipage}[b]{1\linewidth}
\centering{\epsfig{figure=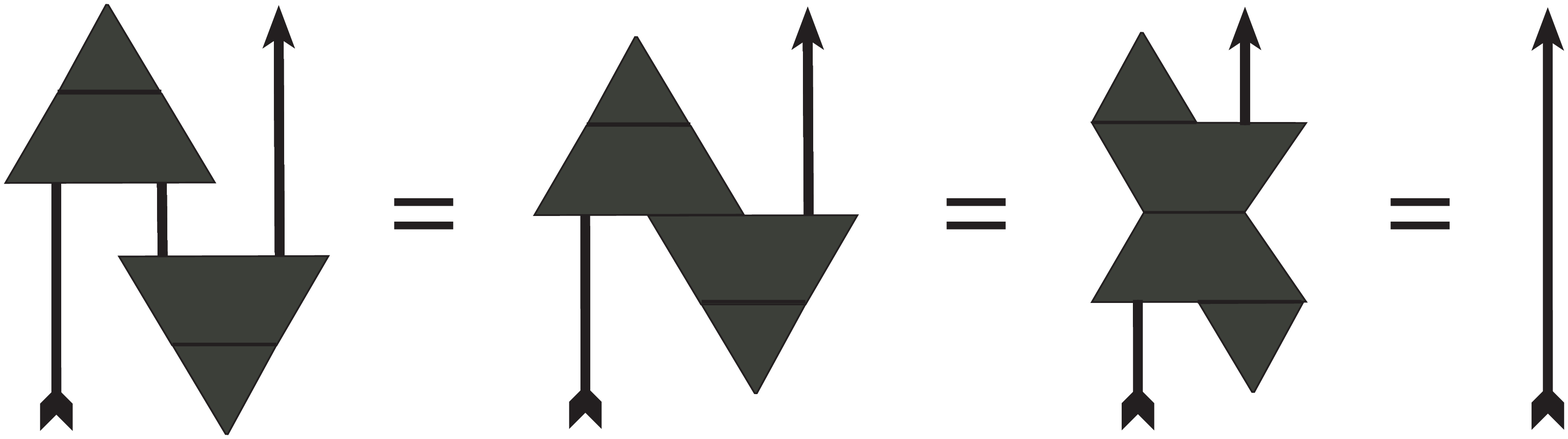,width=252pt}}         
\end{minipage}
\par\vspace{1mm}\par\noindent
that is, $\delta\circ\epsilon^\dagger:\II\to X\otimes X$ and $\epsilon\circ\delta^\dagger:X\otimes X\to\II$ satisfy equations (\ref{eq:compact}) of Section \ref{sec:AQM} and hence canonically provide a unit $\eta=\delta\circ\epsilon^\dagger$ and counit $\varepsilon=\epsilon\circ\delta^\dagger$ which realizes $X^*=X$ (cf.~Section \ref{sec:AQM}).  In a picture this choice stands for:
\par\vspace{2mm}\par\noindent
\begin{minipage}[b]{1\linewidth}
\centering{\epsfig{figure=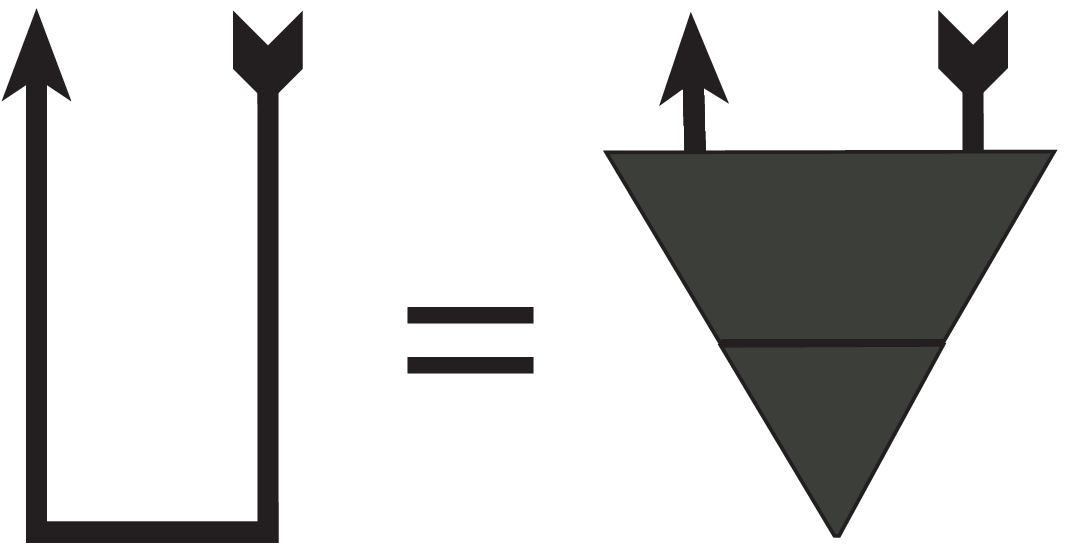,width=83pt}}      
\end{minipage}

One easily verifies that the linear maps $\delta$ and $\epsilon$ as defined in the previous section 
indeed yield an internal comonoid structure on the Hilbert space $\mathbb{C}^{\oplus n}$ which satisfies the \em Frobenius identity \em (\ref{eq:Frobenius}),  and that $\delta\circ\epsilon^\dagger$ is the Bell-state.

\begin{definition}
A \em classical object \em in a $\dagger$-compact category is a special $\dagger$-compact Frobenius algebra $(X,\delta,\epsilon)$ i.e.~a special $\dagger$-Frobenius algebra for which we choose $\eta_X=\delta_X\circ\epsilon_X^\dagger$ in order  to realize $X^*=X$.
\end{definition}
So typical examples of classical objects are the ones existing in {\bf FdHilb} which were implicitly discussed in Section \ref{sec:crux}, namely 
\[
(\mathbb{C}^{\oplus n}, \delta^{(n)}:\mathbb{C}^{\oplus n}\to\mathbb{C}^{\oplus n} \otimes \mathbb{C}^{\oplus n} ::|i\rangle\mapsto|ii\rangle, \epsilon^{(n)}:\mathbb{C}^{\oplus n}\to \mathbb{C}::|i\rangle\mapsto 1).
\]

Since the Frobenius identity (\ref{eq:Frobenius}) allows us to set $X^*=X$ we can now compare $\delta_*, \delta:X\to X\otimes X$,  and also, $\epsilon_*,\epsilon:X\to\II$,  them having the same type.
Recalling that in {\bf FdHilb} the covariant functor $(-)_*$ stands for complex conjugation, the structure of a $\dagger$-compact Frobenius algebra guarantees the highly significant and crucial property that the operations of copying and deleting classical data carry no phase information:
\begin{theorem}\label{thm:phasefree}
For a classical object we have $\delta_*=\delta$ and $\epsilon_*=\epsilon$.
\end{theorem}
Before we prove this fact we introduce some additional concepts.  

\subsection{Self-adjointness relative to a classical object} 

From now on we will denote classical objects as $X$ whenever it is clear from the context that we are considering the structured classical data type $(X,\delta,\epsilon)$ and not the unstructured quantum data type $X$.
Given a classical object $X$ we call a morphism  ${\cal F}:A\to X\otimes A$  \em self-adjoint relative to $X$ \em if the diagram
\beq\label{diag:Xselfadj}
\begin{diagram}
A&\rTo^{\cal F}&X\otimes A\\
\dTo^{\lambda_A}&&\uTo_{1_X\otimes {\cal F}^\dagger}\\
\II\otimes A&\rTo_{\eta_X\otimes 1_A}&X\otimes X\otimes A
\end{diagram}
\eeq
commutes. In a picture, this is:
\par\vspace{2mm}\par\noindent
\begin{minipage}[b]{1\linewidth}
\centering{\epsfig{figure=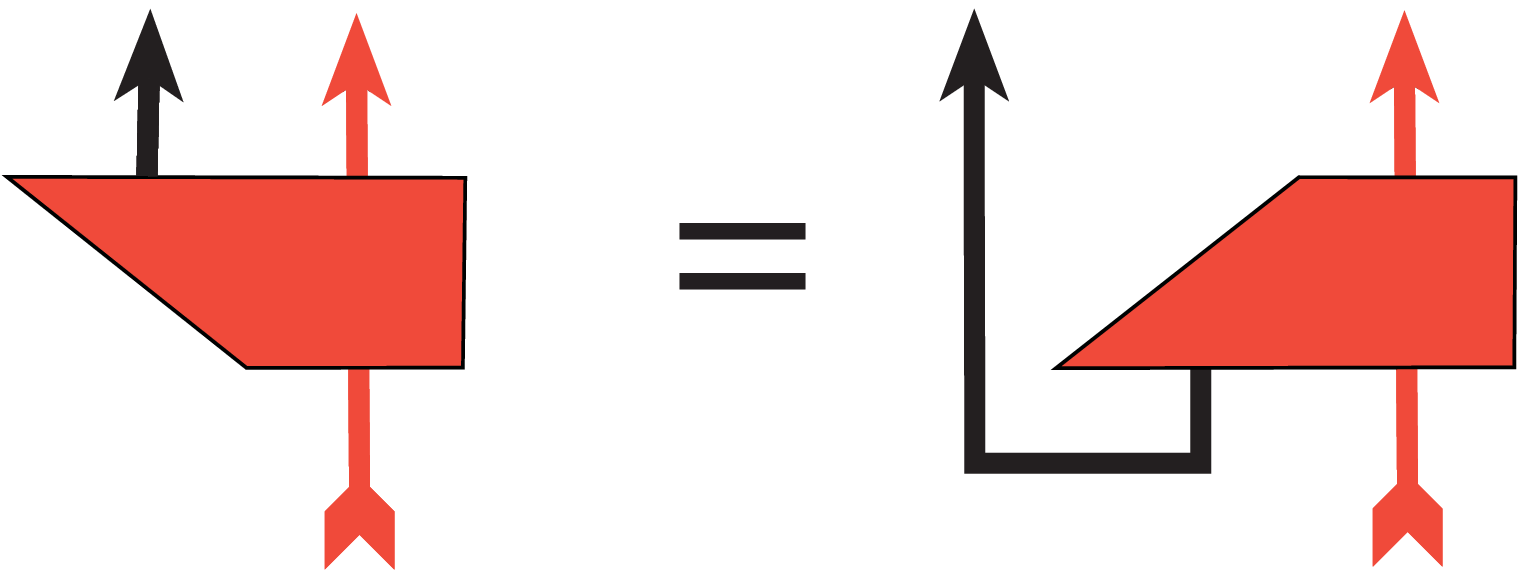,width=290pt}}  
\end{minipage}
A morphism  ${\cal F}:X\otimes A \to A$ is self-adjoint relative to $X$ whenever ${\cal F}^\dagger$ is.  Note furthermore that in every monoidal category, the unit $\II$ carries a canonical comonoid structure, with $\delta = \lambda_\II = \rho_\II:\II\to\II\otimes\II$ and $\epsilon = 1_\II:\II\to \II$. In every $\dagger$-compact category, this comonoid is in fact a degenerate classical object. 
%in {\bf FdHilb}  being 
%\[
%\Bigl(\mathbb{C}\,,\,\lambda_{\mathbb{C}}:\mathbb{C}\to\mathbb{C}\otimes\mathbb{C}::1\mapsto(1,1)\Bigr)\,.
%\]
Self-adjointness in the usual sense of $f^\dagger=f:A\to A$ corresponds to self-adjointness relative to $\II$. For a general classical object $X$, a morphism ${\cal F}:A\to X\otimes A$ can be thought of as an $X$-indexed family of morphisms of type $A\to A$. Self-adjointness relative to $X$ then means that each of the elements of this indexed family are required to be self-adjoint in the ordinary sense. We abbreviate `self-adjoint relative to $X$' to `$X$-self-adjoint'.  There are several analogous generalizations of standard notions e.g.~$X$-scalar, $X$-inverse, $X$-unitarity, $X$-idempotence, $X$-positivity etc.    In Section \ref{sec:coalg}
we discuss a systemic way of defining these `relative to $X$'-concepts. 

\begin{proposition}
Both the comultiplication $\delta$ and the unit $\epsilon$ of a classical object $X$  are  always $X$-self-adjoint, that is, in a picture:
\par\vspace{2mm}\par\noindent
\begin{minipage}[b]{1\linewidth}
\centering{\epsfig{figure=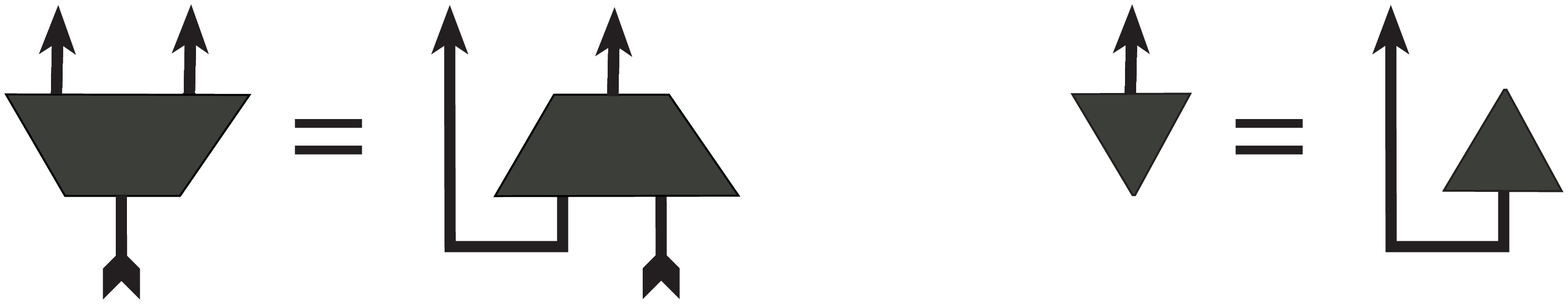,width=240pt}}         
\end{minipage}
\end{proposition} 

\begin{proof}{}
\par\vspace{2mm}\par\noindent
\begin{minipage}[b]{1\linewidth}
\centering{\epsfig{figure=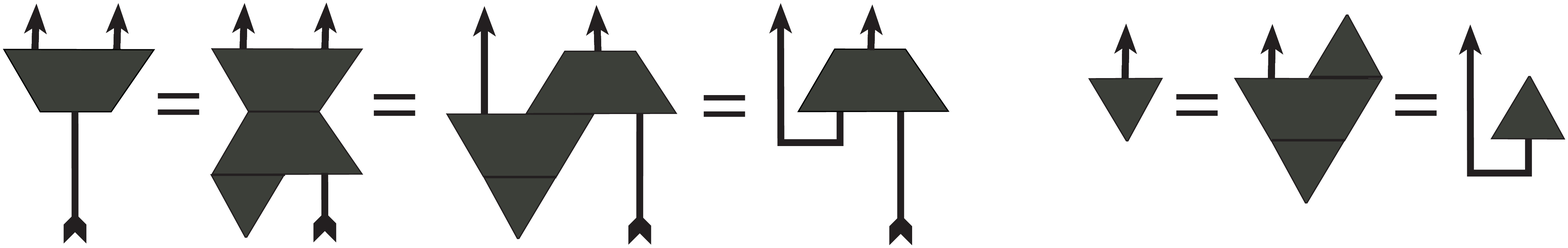,width=280pt}} 
\end{minipage}  
%\par\vspace{-2mm}\par\noindent
\hfill\end{proof}
    
\n Note that $X$-self-adjointness of $\epsilon$ is exactly $\epsilon_*=\epsilon$, already providing part of the proof of Theorem \ref{thm:phasefree}.  In fact, given an internal commutative comonoid $(X,\delta,\epsilon)$ diagram (\ref{diag:Xselfadj}) implicitly stipulates that, of course $X^*=X$, but also that this self-duality of $X$ is realized through $\eta=\delta\circ\epsilon^\dagger$ since we have
\par\vspace{2mm}\par\noindent 
\begin{minipage}[b]{1\linewidth}
\centering{\epsfig{figure=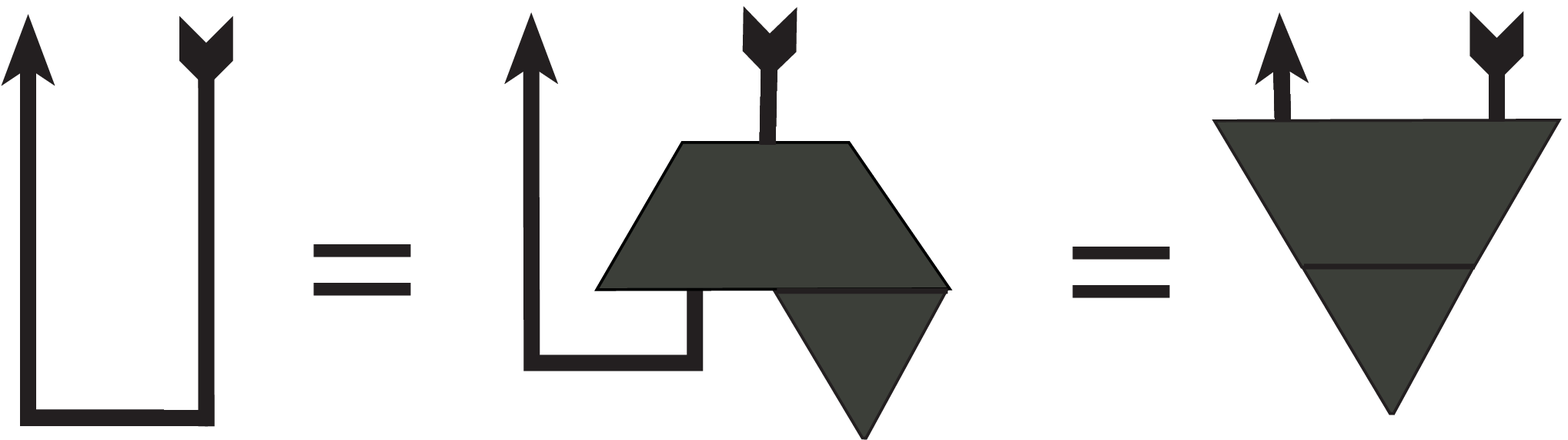,width=166pt}}        
\end{minipage} 
Hence it makes sense to speak of an \em $X$-self-adjoint internal comonoid \em in a $\dagger$-compact category.   From $X$-self-adjointness we can straightforwardly derive many other useful properties, including the Frobenius identity itself, hence providing an alternative characterization of classical objects, and also $\delta_*=\delta$, providing the remainder of the proof of Theorem \ref{thm:phasefree}. 
\begin{lemma}
The comultiplication of an $X$-self-adjoint commutative internal monoid  satisfies the Frobenius identity {\rm(}\ref{eq:Frobenius}{\rm)}, is partial-transpose-invariant\break ${\rm pt}_{\II,X}^{X,X}(\delta) = \delta$, and is self-dual $\delta_*=\delta$ {\rm(}or $\delta^*=\delta^\dagger${\rm)}.  The latter two depict as{\rm\,:}
\par\vspace{2mm}\par\noindent
\begin{minipage}[b]{1\linewidth}
\centering{\epsfig{figure=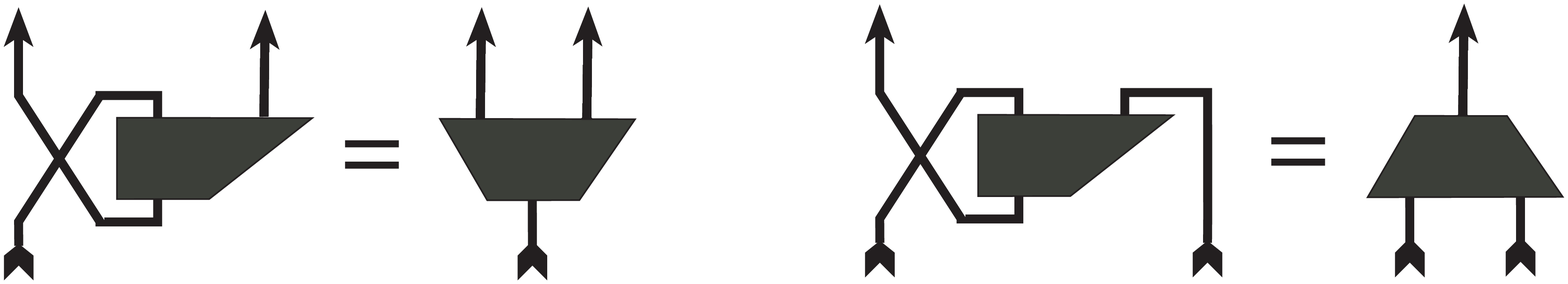,width=281pt}}     
\end{minipage}
\end{lemma}  

\begin{proof}
For the Frobenius identity, apply $X$-self-adjointness to the lefthandside, use associativity of the comultiplication, and apply $X$-self-adjointness again, for 
partial-transpose-invariance apply $X$-self-adjointness twice, and for self-duality apply $X$-self-adjointness three times.
\end{proof}

\begin{theorem}\label{thm:selfSdj}
A classical object can equivalently be defined as a special $X$-self-adjoint internal commutative comonoid $(X,\delta,\epsilon)$. 
\end{theorem}

\subsection{{\sf GHZ}-states as classical objects} 

Analogously to the Hilbert space case (cf.~Section \ref{sec:crux}), each classical object  $X$ induces an abstract counterpart to generalized {\sf GHZ}-states, namely
\[
{\sf GHZ}_X:=(1_X\otimes\delta)\circ\eta:\II\to X\otimes X\otimes X\,.
\]
In a picture that is:
\par\vspace{2mm}\par\noindent
\begin{minipage}[b]{1\linewidth}
\centering{\epsfig{figure=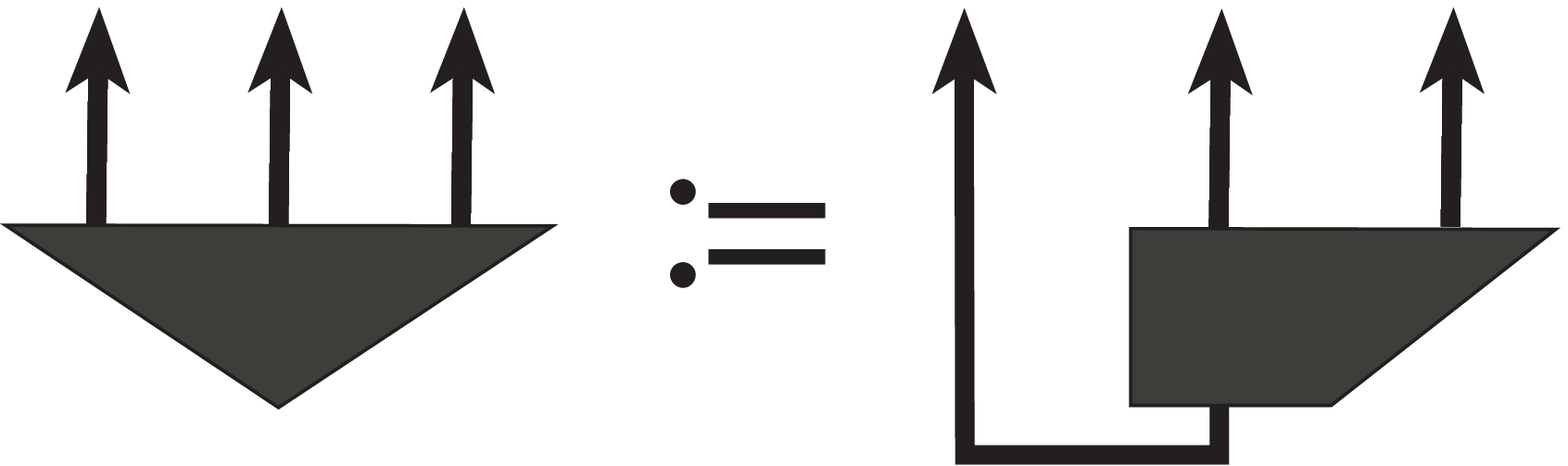,width=126pt}}      
\end{minipage}
The unit property of the comonoid structure, together with the particular choice for the unit of compact closure $\varepsilon=\epsilon\circ\delta^\dagger$ become pleasingly symmetric:
\par\vspace{2mm}\par\noindent
\begin{minipage}[b]{1\linewidth}
\centering{\epsfig{figure=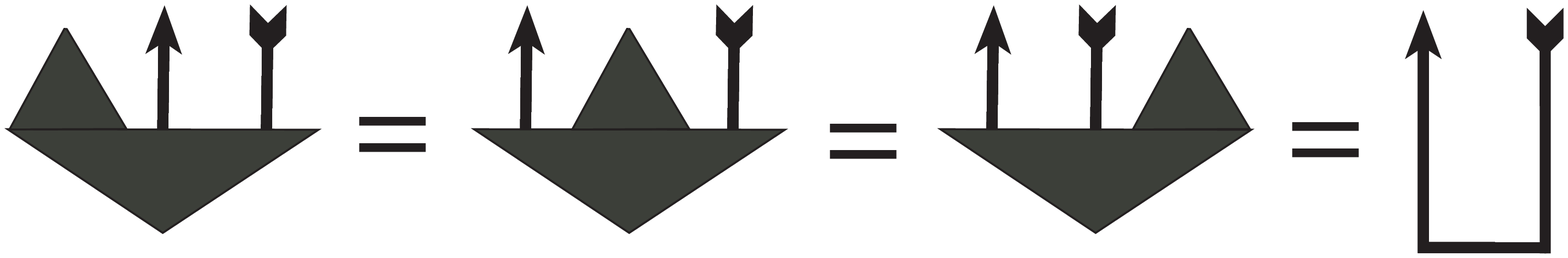,width=223pt}}      
\end{minipage}
\par\vspace{2mm}\par\noindent
The same is the case for commutativity  of the comonoid structure, together with partial-transpose-invariance:
\par\vspace{2mm}\par\noindent
\begin{minipage}[b]{1\linewidth}
\centering{\epsfig{figure=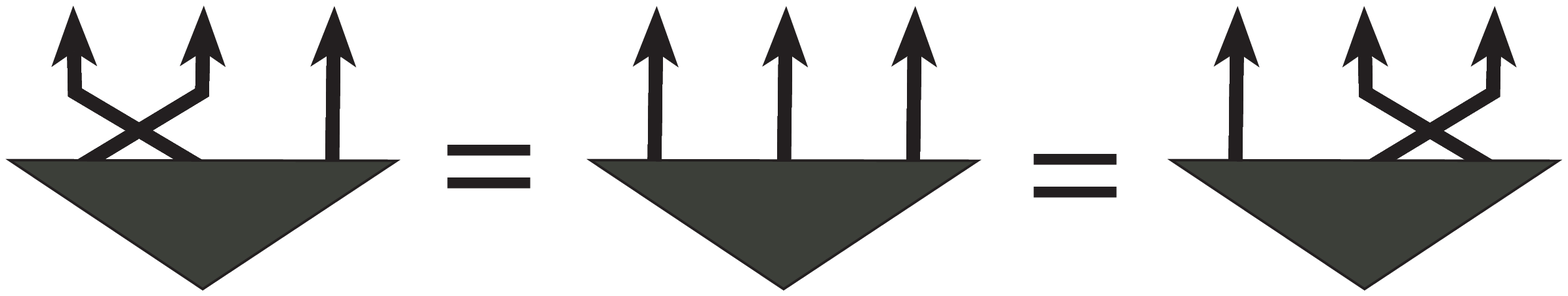,width=185pt}}      
\end{minipage}

\subsection{Extracting the classical world}

If ${\bf C}$ comes with a $\dagger$-structure then any internal comonoid yields an internal monoid. But there is a clear conceptual distinction between the two structures, in the sense that the comultiplication and its unit admit interpretation in terms of copying and deleting.  We will be able to extract the classical world by defining \em morphisms of classical objects \em to be those which preserve the copying and deleting operations of these classical objects, or, in other words, by restricting to those morphisms with respect to which the copying and deleting operations \em become natural \em (cf.~Section \ref{sec:crux}).  

Given a $\dagger$-compact category $\bf C$, we define a new category ${\bf C}_\times$ of which the objects are the classical objects and with the morphisms restricted to those which preserve both $\delta$ and $\epsilon$. So there is a forgetful functor
\[
{\bf C}_\times\ \rightarrow\ {\bf C}\,.
\]
In {\bf FdHilb},  a linear map $f:\mathbb{C}^{\oplus m}\to\mathbb{C}^{\oplus n}$ preserves $\epsilon^{(n)}$ if it is a `pseudo stochastic operator'  i.e.~$\sum_{j=1}^{j=n} f_{ij}= 1$ for all $i$ (note that $f_{ij}$ can still be properly complex), and it preserves  $\delta^{(n)}$ if $f_{ij}f_{ij}=f_{ij}$ and $f_{ij}f_{ik}=0$ for $j\not=k$, hence, there is a function $\varphi:m\to n$ such that 
\[
f(|\,i\rangle) = |\,\varphi(i)\rangle\,. 
\]
So ${\bf FdHilb}_\times={\bf FSet}$, the latter being the category of finite sets and functions.  Hence morphisms in ${\bf C}_\times$ are to be conceived as deterministic manipulations of classical data, i.e.~while ${\bf C}$ represents the quantum world, ${\bf C}_\times$ represent the classical world.   The canonical status of $\bf C_\times$ is exposed by the following result due to Fox  \cite{Fox}.
\begin{theorem} 
Let $\bf C$ be a symmetric monoidal category. The category $\bf C_\times$ of its commutative comonoids and corresponding morphisms, with the forgetful functor $\bf C_\times \to C$, is final among all cartesian categories with a monoidal functor to $\bf C$, mapping the cartesian product $\times$ to the tensor $\otimes$.
\end{theorem}
In fact, there are many other categories of classical operations which can be extracted from ${\bf C}$ using classical object structure, including Carboni and Walters' categories of relations, and categories of (doubly) stochastic maps which, in turns, induce information ordering.  For this we refer the reader to \cite{CPP} and other forthcoming papers.

%!!! note that (-)_* provides the structure of complex conjugation while {\bf FStoch} provides the continuum in terms of probabilities.  The question is whether these two together might provide sufficient structure to describe the whole quantum world.

%\begin{theorem} Let ${\bf C}$ be $\dagger$-compact and let ${\bf C}_\oplus$ be the category of its classical objects, with monoid and comonoid preserving morphisms.  Then ${\bf C}_\oplus$ is $\dagger$-monoidal for the induced tensor structure, which in  ${\bf C}_\oplus$ are  in fact biproducts. \end{theorem}

\section{Quantum spectra}

Given a classical object $X$, a morphism  
${\cal F}:A\to X\otimes A$ is \em idempotent relative to $X$\em, or shorter, \em $X$-idempotent\em, if
\[
\begin{diagram}
A&\rTo^{\cal F}&X\otimes A\\
\dTo^{\cal F}&&\dTo_{1_X\otimes {\cal F}}\\
X\otimes A&\rTo_{\delta\otimes 1_A}&X\otimes X\otimes A
\end{diagram}
\]
commutes. In a picture that is:
\par\vspace{2mm}\par\noindent
\begin{minipage}[b]{1\linewidth}
\centering{\epsfig{figure=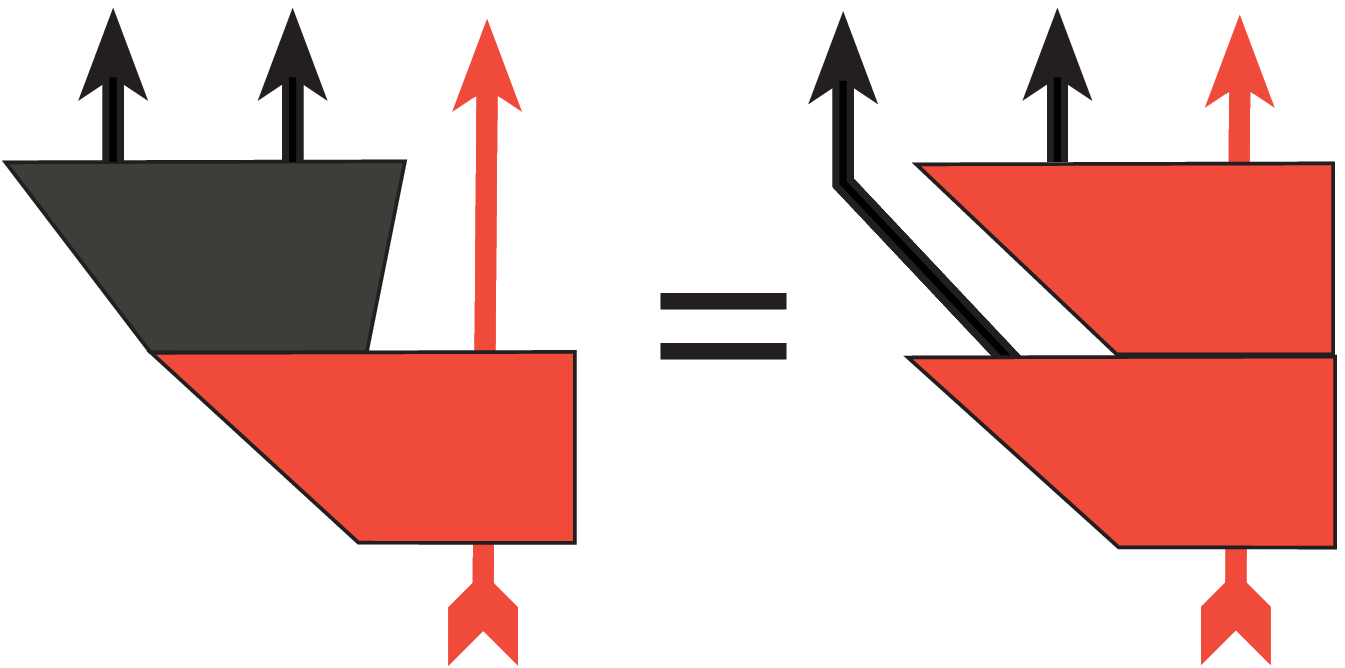,width=113pt}}     
\end{minipage}
\par\vspace{2mm}\par\noindent
Continuing in the same vein, an \em $X$-projector \em is a morphism ${\cal P}:A\to X\otimes A$ which is both $X$-self-adjoint and $X$-idempotent.  The following proposition shows that an $X$-projector is not just an indexed family of projectors.
\begin{proposition}
A $\mathbb{C}^{\oplus k}$-projector in {\bf FdHilb} of type ${\cal H}\to\mathbb{C}^{\oplus k}\otimes{\cal H}$ with ${\cal H}\simeq \mathbb{C}^{\oplus n}$ exactly corresponds to a family of $k$ mutually orthogonal projectors $\{\PP_i\}_{i=1}^{i=k}$, hence we have $\sum_{i=1}^{i=k}\PP_i\leq 1_{\cal H}$.  
\end{proposition}

\begin{proof}{}
One verifies that from $\mathbb{C}^{\oplus k}$-idempotence follows idempotence $\PP_i^2=\PP_i$ and 
mutual orthogonality $\PP_i\circ\PP_{j\not=i}={\bf 0}$, and that from 
$\mathbb{C}^{\oplus k}$-self-adjointness follows orthogonality of projectors $\PP_i^\dagger=\PP_i$.
\end{proof}

\begin{definition}
A morphism ${\cal P}:A\to X\otimes A$  is said to be $X$-complete if  
\[
\lambda_A^\dagger\circ(\epsilon\otimes 1_A)\circ{\cal P}=1_A\,.
\]
In a picture that is:
\par\vspace{2mm}\par\noindent
\begin{minipage}[b]{1\linewidth}
\centering{\epsfig{figure=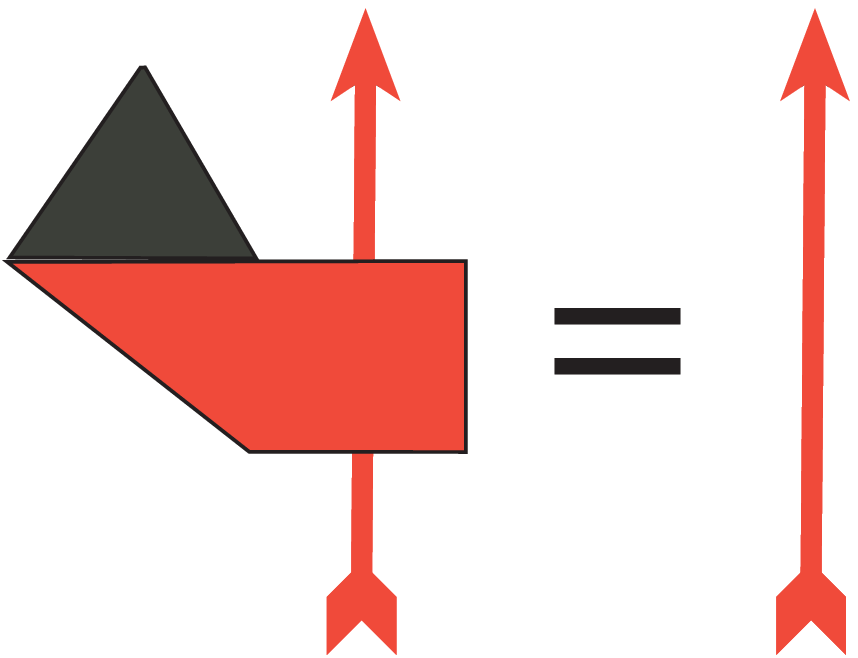,width=71pt}}     
\end{minipage}
\par\vspace{2mm}\par\noindent
A morphism ${\cal P}:A\to X\otimes A$ is a \em projector-valued spectrum \em if it is an $X$-projector for some classical object $X$, and if it is moreover  $X$-complete.
\end{definition}
\begin{theorem}
Projector-valued spectra in {\bf FdHilb} exactly correspond to complete families of mutually orthogonal projectors $\{\PP_i\}_i$, i.e.~$\sum_{i=1}^{i=k}\PP_i= 1_{\cal H}$.
\end{theorem}
Each classical object $(X,\delta,\epsilon)$ canonically induces a projector-valued spectrum $\delta:X\to X\otimes X$
since associativity of the comultiplication coincides with $X$-idempotence and  the defining property of the comultiplicative unit coincides with completeness --- the reader should not be confused by the fact that the quantum data type $X$ is now also the carrier of the classical data type $(X,\delta,\epsilon)$.  Having in mind the characterization of classical objects of Theorem \ref{thm:selfSdj}, mathematically, projector-valued spectra consitute a generalization of classical objects by admitting \em degeneracy\em.

\subsection{Coalgebraic characterization of spectra}\label{sec:coalg} 

Recall  from \cite{Dusko} that the internal commutative (co)monoid structures over an object $X$ in a monoidal category ${\bf C}$ are in one-to-one correspondence with commutative (co)monad structures on the functor 
\[
X\otimes-:{\bf C}\to{\bf C}\,. 
\]
Hence we can attribute a notion of (co)algebra to internal commutative (co)monoids.
\begin{theorem}
Let ${\bf C}$ be a $\dagger$-compact category. Its projector-valued spectra are exactly  the $X$-self-adjoint Eilenberg-Moore coalgebras for the comonads $X\otimes-:{\bf C}\to{\bf C}$ canonically induced by some classical object $X$.
\end{theorem}

\begin{proof}{}
The requirements for Eilenberg-Moore coalgebras with respect to the comonad ${(X\otimes-)}$ are exactly $X$-idempotence and $X$-complete\-ness.
\end{proof}
We can now rephrase all the above as follows.
\begin{theorem}
$X$-self-adjoint coalgebras in {\bf FdHilb} exactly correspond to complete families of mutually orthogonal projectors $\{\PP_i\}_i$.
\end{theorem} 

\begin{proof}
We also rephrase the proof.
From the Eilenberg-Moore commuting square  we obtain idempotence $\PP_i^2=\PP_i$ and 
mutual orthogonality $\PP_i\circ\PP_{j\not=i}={\bf 0}$, from the Eilenberg-Moore commuting triangle we obtain completeness and from $X$-self-adjointness follows orthogonality of projectors $\PP_i^\dagger=\PP_i$.
\end{proof}

\subsection{Characterization of $X$-concepts} 

All `relative to $X$'-concepts can now be defined  as the corresponding standard concept in the Kleisli category for the comonad ${(X\otimes-)}$.  For example, $X$-unitarity of a morphism ${\cal U}:X\otimes A\to A$ simply means that ${\cal U}$ is unitary in the Kleisli category for the comonad ${(X\otimes-)}$.
This approach immediately provides all coherence conditions which are required for these $X$-concepts to be sound with respect to the categorical structure with which one works.   Below we also define $X$-unitarity in an ad hoc manner for those readers which are not very familiar yet with categorical language.

\section{Quantum measurements}

Given projector-valued spectra we are very close to having an abstract notion of quantum measurement. In fact, the type $A\to X\otimes A$ which we attributed to the spectra is indeed the compositional type of a (\em non-demolition\em) measurement.  But what is even more compelling is the following.  The fact that a spectum is  $X$-idempotent, or equivalently, that it satisfies the coalgebraic Eilenberg-Moore commuting square, i.e. 
\begin{diagram} 
A&\rTo^{Measure}&X\otimes A\\
\dTo^{\qquad Measure}&&\dTo_{1_X\otimes Measure}\\
X\otimes A&\rTo_{Copy\otimes 1_A}&X\otimes X\otimes A
\end{diagram}
exactly captures  \em von Neumann's projection postulate\em, stating that repeating a measurement is equivalent to copying the data obtained in its first execution. Note here in particular the manifest \em resource sensitivity \em of this statement, accounting for the fact that two measurements provide two sets of data, even if this data turns out to be identical.  

However, what we get in {\bf FdHilb} is not (yet) a quantum measurement. For $A=X:=\mathbb{C}^{\oplus n}$ the canonical projector-valued spectrum $\delta^{(n)}:A\to X\otimes A$ expressed in the computational base yields
\begin{diagram}
\hspace{-3.5cm}|\,\psi\rangle=\sum_i\langle\, i\,|\,\psi\rangle|\,i\rangle_A&\rMo&\sum_i\langle\, i\,|\,\psi\rangle\left(|\,i\,\rangle_X\otimes|\,i\,\rangle_A\right)\hspace{-4cm}
\end{diagram}
where $|\,i\,\rangle_X\in X$ is the measurement outcome, $|\,i\,\rangle_A\in A$ is the resulting state of the system for that outcome,  and the coefficients $\langle\, i\,|\,\psi\rangle$ in the sum capture the respective probabilities for these outcomes i.e.~$|\langle\, i\,|\,\psi\rangle|^2$. 
This however does not reflect the fact that we cannot retain the relative phase factors present in the probability amplitudes $\langle\, i\,|\,\psi\rangle$.  In other words, the passage from physics to the semantics is not \em fully abstract\em.  
%\footnote{Formally, full abstraction is a notion introduced by Abramsky in \cite{FullyAbstract}.} 
It is moreover well-known that the operation which erases these relative phases does not live in {\bf FdHilb}, but is \em quadratic \em in the state, hence lives in ${\bf CPM}({\bf FdHilb})$, the category of Hilbert spaces and completely positive maps. 

Fortunately, for many practical purposes (such as those outlined in Sections \ref{teleportation} and \ref{densecoding} of this paper) this `approximate' notion of measurement suffices,\footnote{This approximate notion of quantum measurement is also the one considered in \cite{AC1}.} and in all other cases it turns out that we can rely on  
Selinger's abstract counterpart for the passage from {\bf FdHilb} to  ${\bf CPM}({\bf FdHilb})$, a construction which applies to any $\dagger$-compact category \cite{Selinger2}, to turn those approximate quantum measurements into true quantum measurements. 
 
\subsection{The ${\bf CPM}$-construction} 

This construction takes a $\dagger$-compact category ${\bf C}$ as its input and produces an `almost inclusion' (it in fact kills redundant global phases) of ${\bf C}$ into a bigger one ${\bf CPM}({\bf C})$. While  ${\bf C}$ is to be conceived as containing pure operations with those of type $\II\to A$ being the pure states, ${\bf CPM}({\bf FdHilb})$ consists of mixed operations with those of type $\II\to A$ being the mixed states. Explicitly we have the $\dagger$-compact functor 
\[
{\sf Pure}:{\bf C}\to {\bf CPM}({\bf C})::f\to f\otimes f_*\,. 
\]
where
\[
{\bf CPM}({\bf C})(A,B):=\Bigl\{
(1_B\otimes \eta^\dagger_{C^*}\otimes 1_{B^*})\circ (f\otimes f_*)
\Bigm|
f:A\to B\otimes C
\Bigr\}
\]
and the $\dagger$-compact structure on  ${\bf CPM}({\bf C})$ covariantly inherits its composition, its tensor, its adjoints and its Bell-states from {\bf C}.  In a picture the morphisms of ${\bf CPM}({\bf C})$  
are:
\par\vspace{2mm}\par\noindent
\begin{minipage}[b]{1\linewidth}
\centering{\epsfig{figure=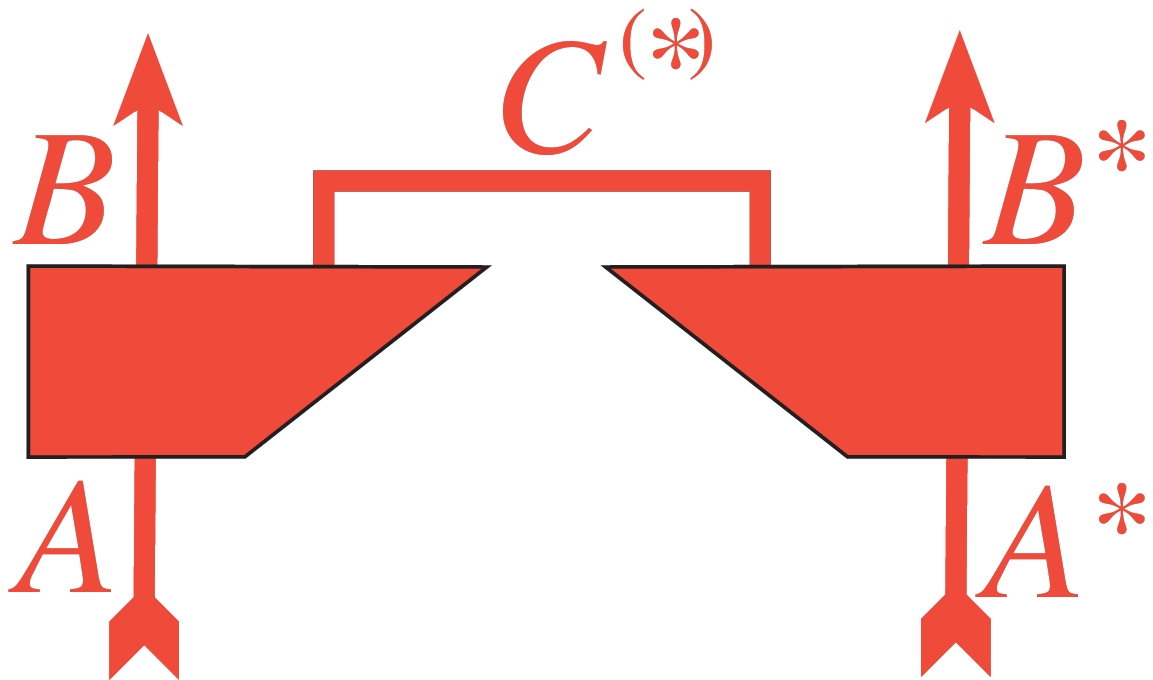,width=90pt}}        
\end{minipage}
\par\vspace{2mm}\par\noindent
Note in particular that the two copies of each ${\bf C}$-morphism in these ${\bf CPM}({\bf C})$-morphisms is also present in  Dirac's notation when working with density matrices. However, in Dirac notation one considers the pair of a \em ket\em-vector $|\,\psi\rangle$ and its adjoint $\langle\psi\,|$ resulting in the action of an operation being 
\[
 |\,\psi\rangle\langle\psi\,|\mapsto f |\,\psi\rangle\langle\psi\,| f^\dagger
\]
for an ordinary operation, while it becomes
\[
 |\,\psi\rangle\langle\psi\,|\mapsto f (1_C\otimes |\,\psi\rangle\langle\psi\,| )f^\dagger
\]
for a completely positive map.  What we do here is quite similar but now we consider pairs  $|\,\psi\rangle\otimes|\,\psi\rangle_*$ allowing for more intuitive covariant composition 
\[
|\,\psi\rangle\otimes|\,\psi\rangle_*\mapsto (f\otimes f_*)(|\,\psi\rangle\otimes|\,\psi\rangle_*)
\]
for an ordinary operation, while it becomes
\[
|\,\psi\rangle\otimes|\,\psi\rangle_*\mapsto (1_B\otimes\eta_C^\dagger\otimes 1_{B^*})(f\otimes f_*)(|\,\psi\rangle\otimes|\,\psi\rangle_*)
\]
for a completely positive map.  The most important benefit of this covariance is 2-dimensional display-ability i.e.~it enables graphical calculus. 
%A genuine example of a properly mixed states in  ${\bf CPM}({\bf C})$  are the Bell-states of ${\bf C}$ corresponding to the \em maximally mixed state\em, which we will denote by $\bot_A:\II\to A$.   In terms of Dirac notation, this 

\subsection{Formal decoherence} 

Given a classical object $X$ in a $\dagger$-compact category ${\bf C}$ we define the following morphism
\[
\Gamma_X:=(1_X\otimes \eta^\dagger\otimes 1_X)\circ (\delta\otimes\delta)\in {\bf CPM}({\bf C})(X,X)\,.
\]
In a picture that is:  
\par\vspace{2mm}\par\noindent
\begin{minipage}[b]{1\linewidth}
\centering{\epsfig{figure=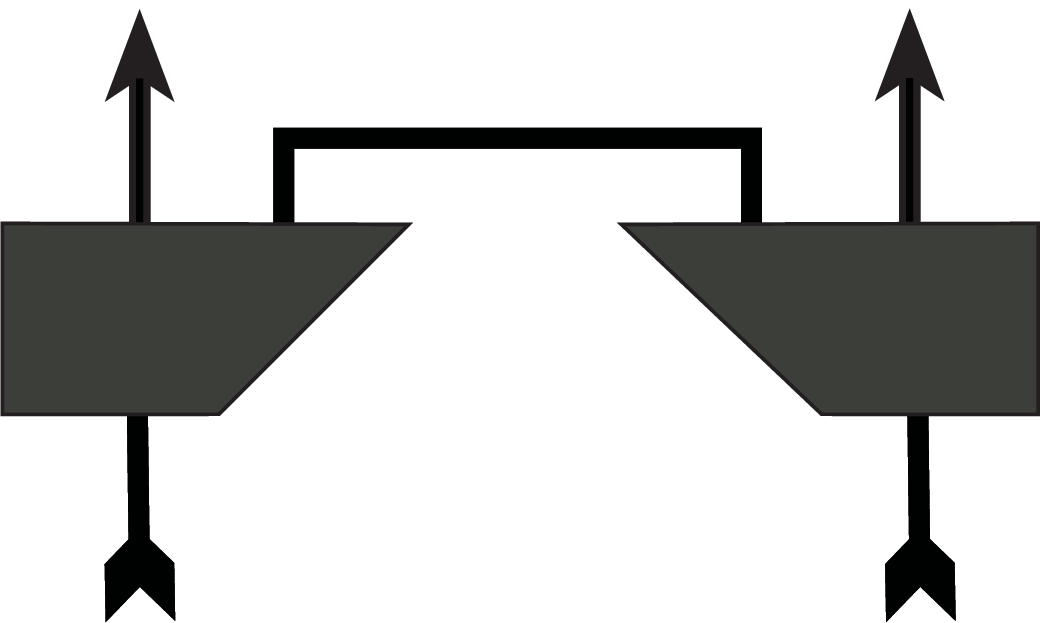,width=90pt}}         
\end{minipage}
\par\vspace{2mm}\par\noindent
\begin{proposition}
In $\dagger$-compact category with $X$ a classical object we have
\[
\Gamma_X=\delta\circ \delta^\dagger:X\otimes X\to X\otimes X
\]
so in particular is $\Gamma_X$ idempotent. 
\end{proposition}

\begin{proof}{}
Using the Frobenius identity we have 
\par\vspace{2mm}\par\noindent
\begin{minipage}[b]{1\linewidth}
\centering{\epsfig{figure=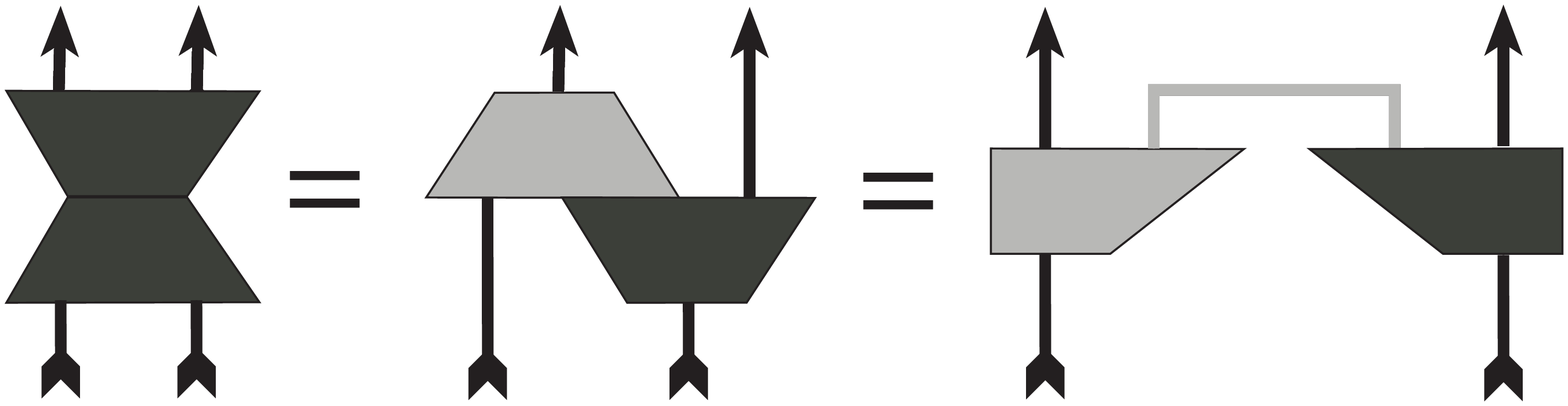,width=190pt}}        
\end{minipage}
\par\vspace{0mm}\par\noindent 
where the highlighted part expresses the use of $X$-self-adjointness.
\end{proof}
In particular in {\bf FdHilb} we have 
\begin{diagram}
\sum_{ij}\alpha_i\bar{\alpha}_j |i\rangle\otimes|j\rangle_*\\
\dMo~{\delta^{(k)}\otimes\delta^{(k)}}&&&&\\
\sum_{ij}\alpha_i\bar{\alpha}_j |ii\rangle\otimes|jj\rangle_*&&\rdTo^{\Gamma_{\mathbb{C}^{\oplus n}}}&&\\
\dMo~{1_{\mathbb{C}^{\oplus n}}\otimes \eta^\dagger_{\mathbb{C}^{\oplus n}}\otimes 1_{\mathbb{C}^{\oplus n}}}&&&&\\
\sum_{ij}\delta_{ij}\alpha_i\bar{\alpha}_j |i\rangle\otimes|j\rangle_*&&\rIs&&\sum_i\alpha_i\bar{\alpha}_i|i\rangle\otimes|i\rangle_*
\end{diagram} 
i.e.~we obtain the desired effect of elimination of the relative phases.  Hence, given a projector-valued spectrum now represented in ${\bf CPM}({\bf C})$ through the functor ${\sf Pure}$, which depicts as
\par\vspace{2mm}\par\noindent
\begin{minipage}[b]{1\linewidth}
\centering{\epsfig{figure=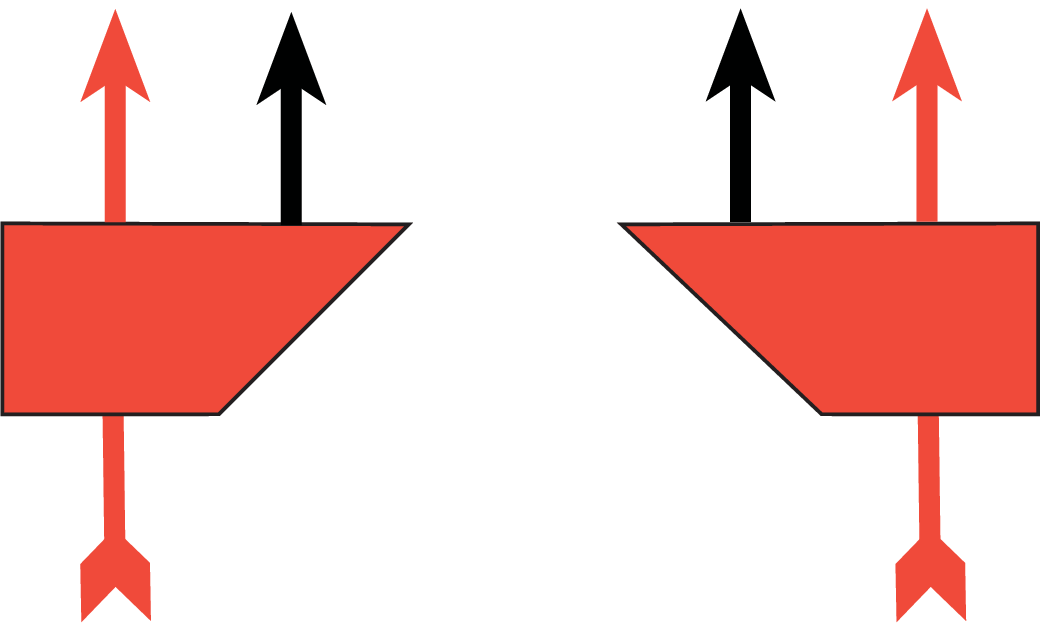,width=85pt}}        
\end{minipage}
\par\vspace{2mm}\par\noindent
we obtain a genuine quantum measurement by adjoining $\Gamma_X$ as in
\[
Meas:=(1_B\otimes\Gamma_X\otimes 1_B)\circ({\cal M}\otimes{\cal M}_*)\,,
\]
which in a picture becomes:
\par\vspace{2mm}\par\noindent
\begin{minipage}[b]{1\linewidth}
\centering{\epsfig{figure=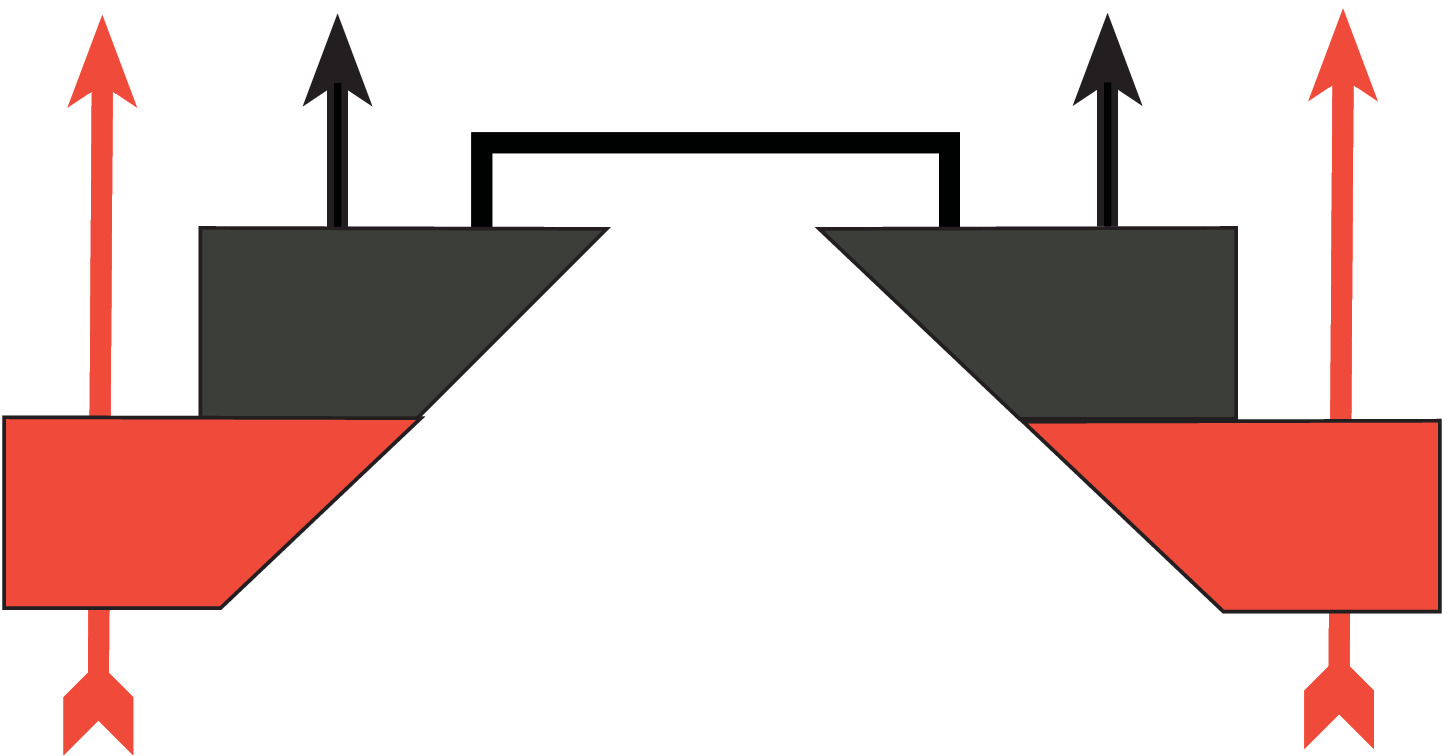,width=115pt}}         
\end{minipage}

\subsection{Demolition measurements} 

As compared to the type $A\to X\otimes A$ of a non-demolition measurement, a \em demolition measurement \em has type $A\to X$.  We claim that the demolition analogue to a projector-valued spectrum ${\cal M}:A\to X\otimes A$ is the adjoint to an \em isometry \em $m^\dagger:X\to A$, i.e.~$m\circ m^\dagger=1_X$ --- or equivalently put in our $X$-jargon, a \em normalized $X$-bra\em. 
%Indeed, since a projector $\PP$ is both idempotent and self-adjoint, it is also \em positive\em, that is, can be written as $f^\dagger\circ f$ for some morphism, with in the case of a projector $f:=\PP$. 
Indeed, setting 
\[
{\cal M}_m:=(1_X\otimes m^\dagger)\circ \delta\circ m:A\to X\otimes A
\]
we exactly obtain a projector-valued spectrum since ${\cal M}_m$ is trivially $X$-self-adjoint, and $m\circ m^\dagger=1_X$ yields  $X$-idempotence. In a picture ${\cal M}_m$ is:
\par\vspace{2mm}\par\noindent
\begin{minipage}[b]{1\linewidth}
\centering{\epsfig{figure=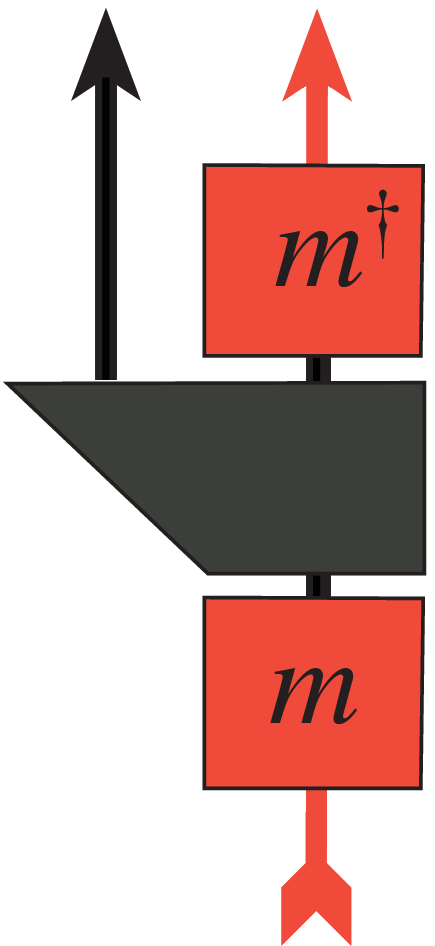,width=37pt}}       
\end{minipage}
\par\vspace{2mm}\par\noindent
The corresponding demolition measurement arises by adjoining $\Gamma_X$ i.e.
\[
DeMeas:=\Gamma_X\circ(m\otimes m_*)\,,
\]
that is, in a picture:
\par\vspace{2mm}\par\noindent
\begin{minipage}[b]{1\linewidth}
\centering{\epsfig{figure=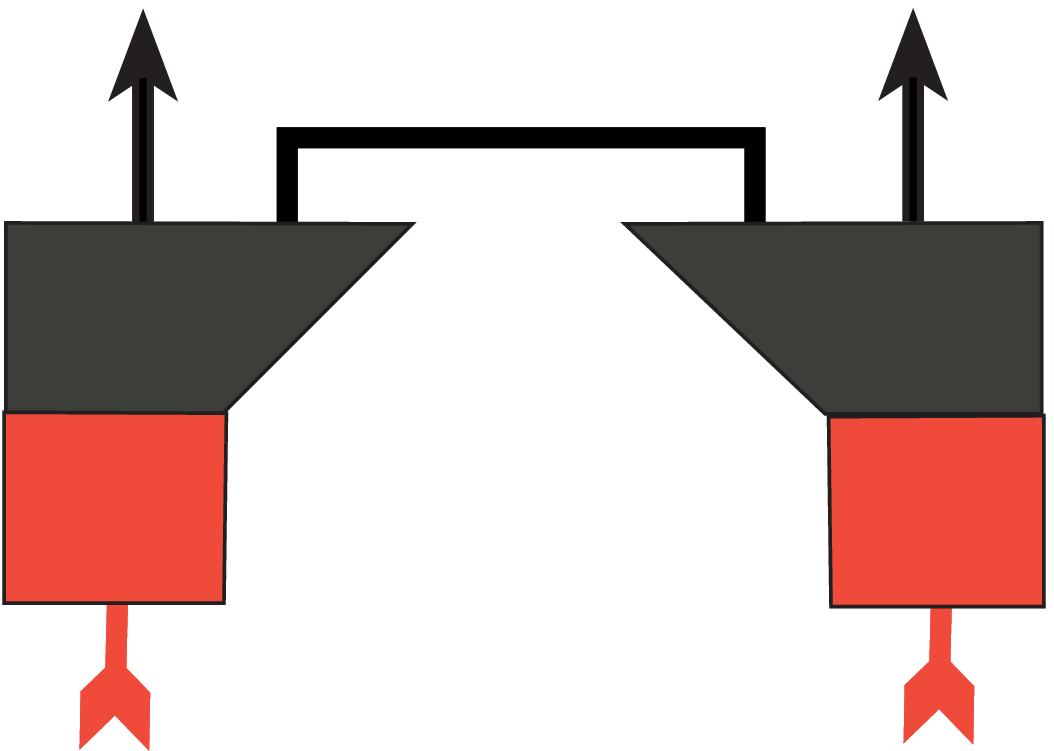,width=85pt}}         
\end{minipage}
Such a demolition measurement is \em non-degenerate \em iff $m$ is unitary.

\section{Quantum teleportation}\label{teleportation}

The notion of measurements proposed in this paper abstracts over the structure of classical data, and we will show that we can describe and proof correctness of the teleportation protocol without making the classical data structure explicit, nor by relying on the cartesian structure of ${\bf C}_\times$.    
\begin{definition}
Given a classical object $X$ a morphism ${\cal U}:X\otimes A\to B$, and at the same time  ${\cal U}^\dagger$ and ${\cal U}\circ\sigma_{X,A}$, are \em unitary relative to $X$ \em or \em $X$-unitary \em iff
\[
(1_X\otimes {\cal U})\circ(\delta\otimes 1_A):X\otimes A\to X\otimes B  
\]
is unitary in the usual sense i.e.~its adjoint is its inverse. 
In a picture:
\par\vspace{2mm}\par\noindent
\begin{minipage}[b]{1\linewidth}
\centering{\epsfig{figure=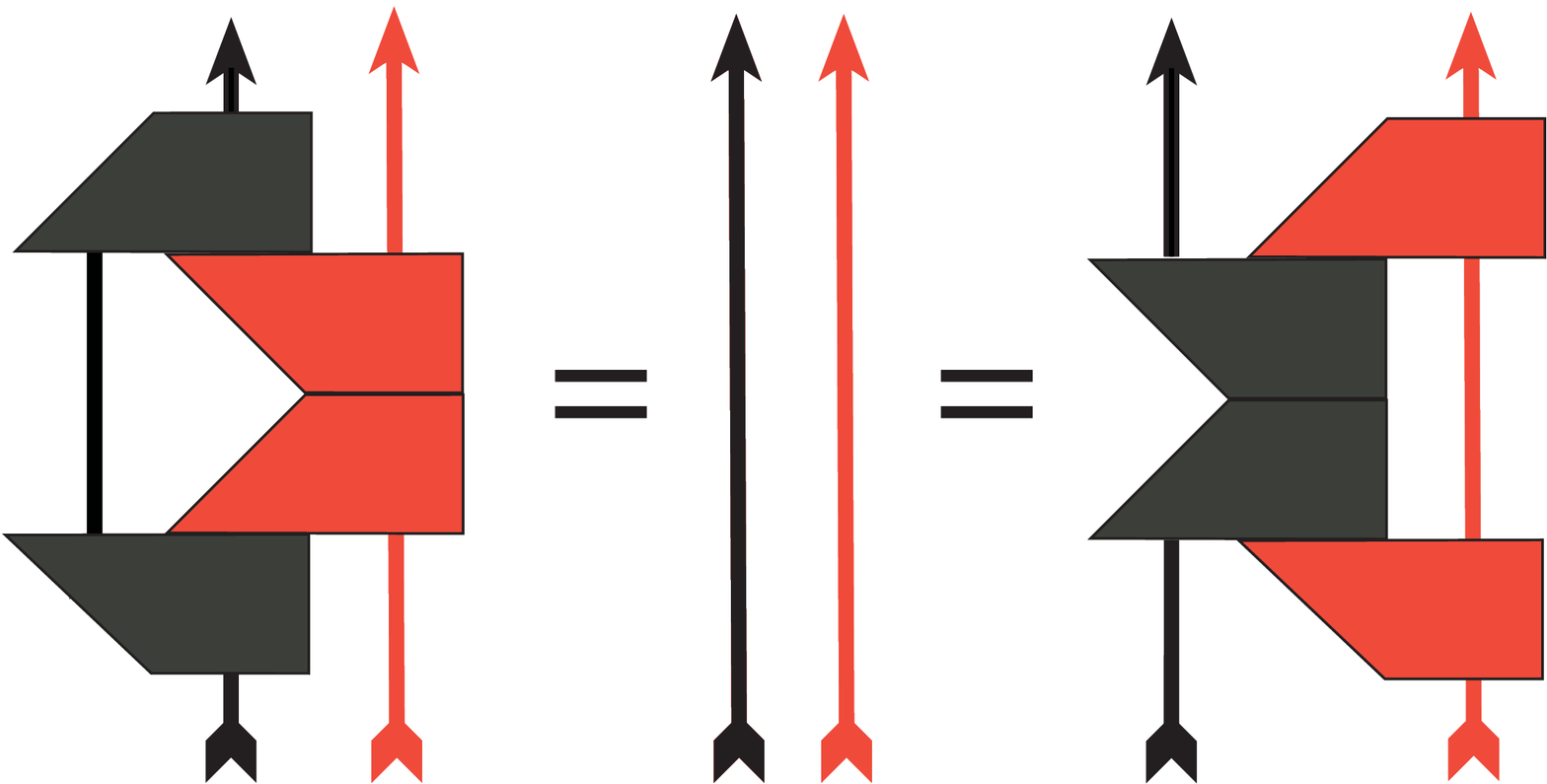,width=155pt}}        
\end{minipage}
\par\vspace{2mm}\par\noindent
\end{definition}
A trivial example of such a unitary morphism is $\epsilon\otimes 1_A:X\otimes A\to A$.  
\begin{comment}%%%%comment:
One easily verifies that the above graphically depicted condition is the same as:
\par\vspace{2mm}\par\noindent
\begin{minipage}[b]{1\linewidth}
\centering{\epsfig{figure=FLOPS14.pdf,width=155pt}}        
\end{minipage}
\par\vspace{2mm}\par\noindent
\end{comment}%%%%endcomment
\begin{proposition}
In {\bf FdHilb} morphisms that are $(\mathbb{C}^{\oplus n},\delta^{(n)})$-unitary are in bijective correspondence with $n$-tuples unitary operators of the same type.
\end{proposition}

\medskip\noindent  
Let the \em size of a classical object \em be the scalar 
\[
s_X:=\eta^\dagger_X\circ\eta_X=\epsilon_X\circ\epsilon^\dagger_X:\II\to\II
\]
i.e.~in a picture:
\par\vspace{2mm}\par\noindent
\begin{minipage}[b]{1\linewidth}
\centering{\epsfig{figure=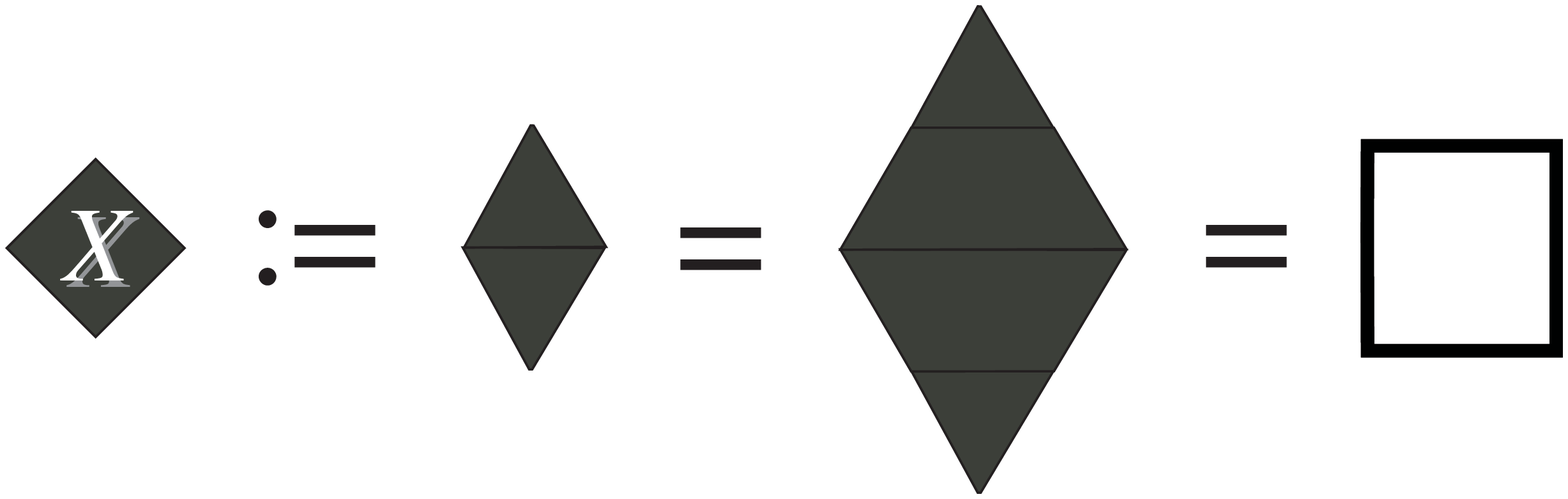,width=170pt}}        
\end{minipage}
\par\vspace{2mm}\par\noindent
using in the last two steps respectively $\delta^\dagger\circ\delta=1_X$ and  $\eta=\delta\circ\epsilon^\dagger$.

\begin{proposition} 
The positive scalars in the scalar monoid ${\bf C}(\II,\II)$, i.e.~those scalars $s:\II\to\II$ that can be written as $s=\psi^\dagger\circ\psi$ for some $\psi:\II\to A$, have self-adjoint square-roots when embedded in  ${\bf CPM}({\bf C})$ via {\sf Pure}.
\end{proposition}

\begin{proof}{}
The image of a positive scalar $s$ under ${\sf Pure}$  is $s\otimes s_*$. For
$t=\eta_{A^*}^\dagger\circ(\psi\otimes\psi_*)\in {\bf CPM}({\bf C})(\II,\II)$
which we depict in a picture as:
\par\vspace{2mm}\par\noindent
\begin{minipage}[b]{1\linewidth}
\centering{\epsfig{figure=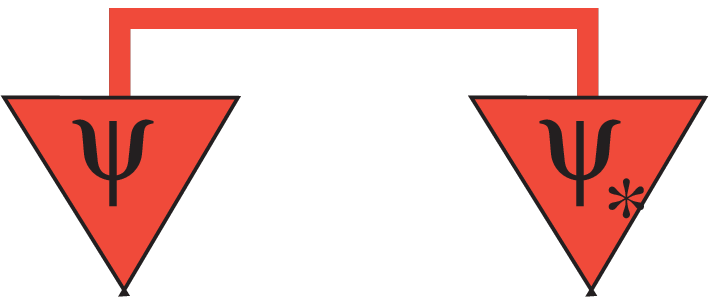,width=76pt}}     
\end{minipage}
\par\vspace{2mm}\par\noindent
we have $t\circ t=s\otimes s_*$ since 
\par\vspace{2mm}\par\noindent
\begin{minipage}[b]{1\linewidth}
\centering{\epsfig{figure=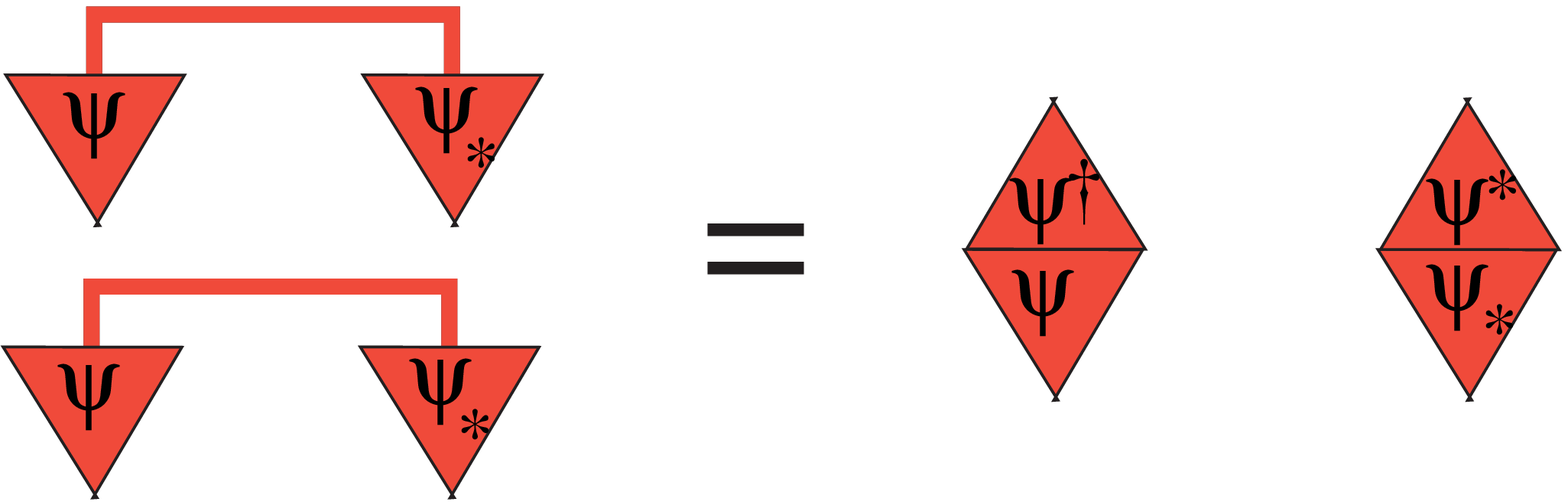,width=221pt}}     
\end{minipage}
\par\vspace{2mm}\par\noindent
follows from $(f^*\otimes 1_B)\circ\eta_B=(1_A\otimes f)\circ\eta_A$ \cite{AC1}. Self-adjointness follows from:
\par\vspace{2mm}\par\noindent
\begin{minipage}[b]{1\linewidth}
\centering{\epsfig{figure=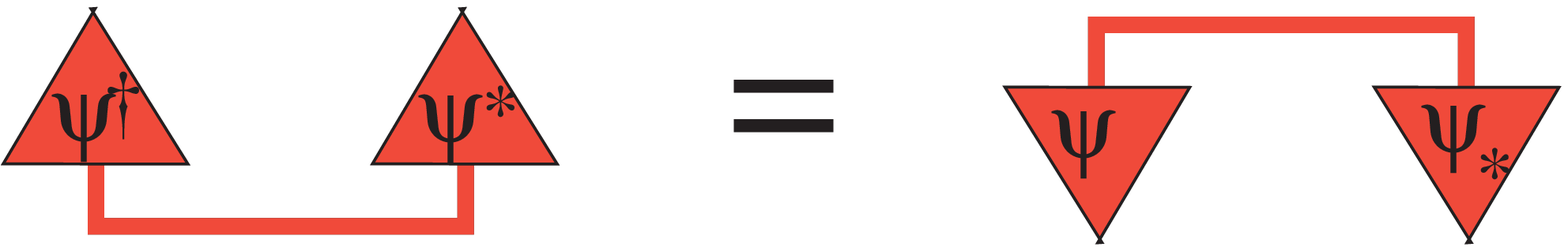,width=221pt}}     
\end{minipage}
\par\vspace{2mm}\par\noindent
Hence ${\sf Pure}(s)$ indeed has a scalar in ${\bf CPM}({\bf C})(\II,\II)$ as a square-root. 
\end{proof} 

\noindent This implies that square-root $\sqrt{s_A}:\II\to\II$ of the  \em dimension \em $s_A:=\eta_A^\dagger\circ\eta_A$ of an object $A$ always exist whenever we are within ${\bf CPM}({\bf C})$.  It can be shown that each $\dagger$-compact category also admits a canonical embedding
in another $\dagger$-compact category in which all scalars have inverses.  For scalars
\[
\ \ s_A\qquad\sqrt{s_A}\qquad\ {1\over s_A}\qquad{1\over\sqrt{s_A}} 
\]
respectively we  introduce the following graphical notations:
\par\vspace{2mm}\par\noindent
\begin{minipage}[b]{1\linewidth}
\centering{\epsfig{figure=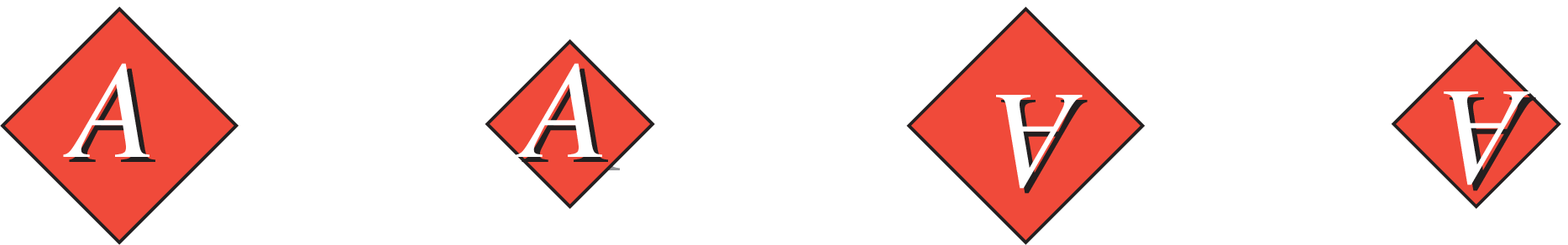,width=150pt}}      
\end{minipage}
\par\vspace{2mm}\par\noindent
--- the reversed symbols representing inverses needn't be confused with the adjoint since these scalar dimensions are always self-adjoint.

\begin{definition}
Let $X$ be a classical object in a $\dagger$-compact category.
A (non-degenerate) \em demolition Bell-measurement  \em is  a unitary morphism 
\[
DeMeas_{Bell}:={1\over\sqrt{s_A}}\bullet\rho_A^\dagger\circ(1_X\otimes \eta^\dagger_A )\circ({\cal U}^\dagger\otimes1_A):A\otimes A^*\to X
\]
which is such that  ${\cal U}:X\otimes A\to A$ is $X$-unitary.
\end{definition}
The corresponding projector-valued spectrum is
\[
{\cal M}_{Bell}:=(DeMeas_{Bell}^\dagger\otimes 1_X)\circ\delta\circ DeMeas_{Bell}:A\otimes A^*\to X\otimes A\otimes A^*\,,
\]
from which the corresponding non-demolition Bell-measurement arises by adjoining $\Gamma_X$.
In a picture the demolition Bell-measurement and corresponding projector-valued spectrum are:
\par\vspace{2mm}\par\noindent
\begin{minipage}[b]{1\linewidth}
\centering{\epsfig{figure=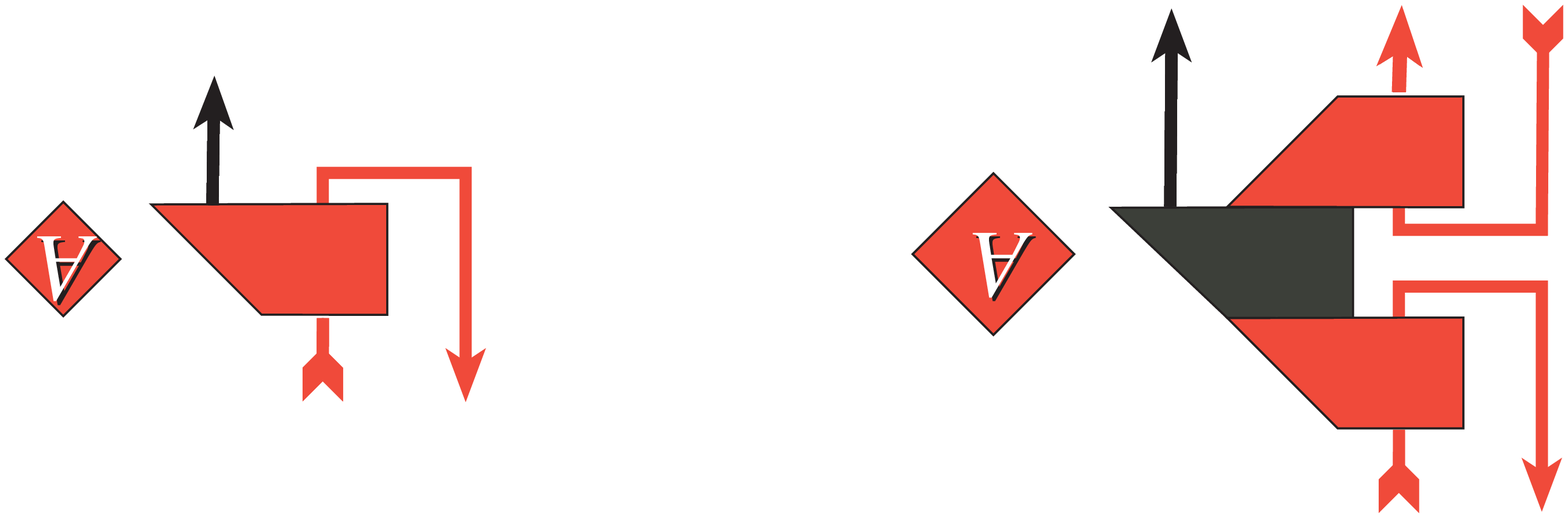,width=180pt}}        
\end{minipage}
\par\vspace{2mm}\par\noindent
and unitarity of $DeMeas_{Bell}$ is:
\par\vspace{2mm}\par\noindent
\begin{minipage}[b]{1\linewidth}
\centering{\epsfig{figure=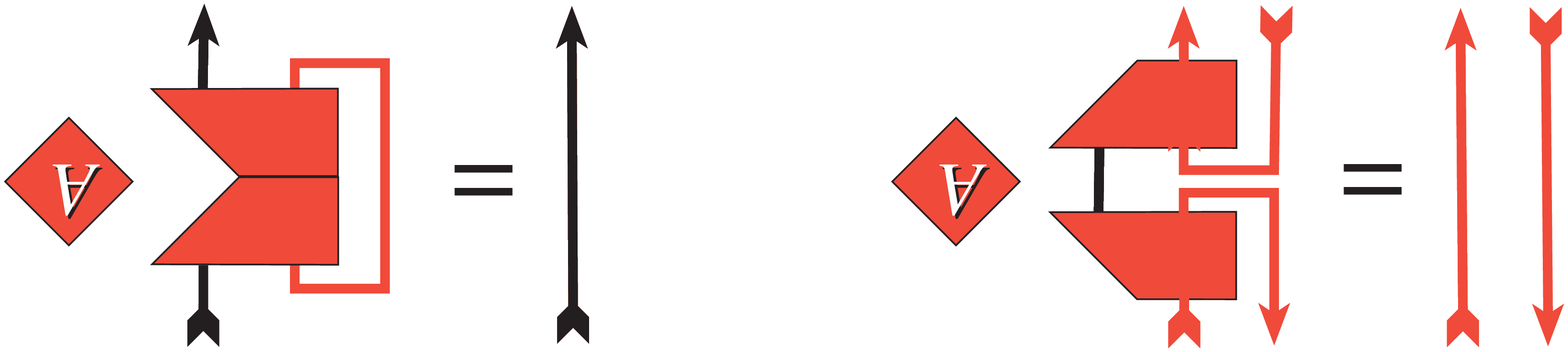,width=240pt}}        
\end{minipage}
\par\vspace{2mm}\par\noindent
that is, in formulae, respectively,
\beq\label{traceunitarity}
{1\over\sqrt{s_A}}\bullet{\rm tr}^A_{X,X}({\cal U}^\dagger\circ{\cal U})=DeMeas_{Bell}\circ DeMeas_{Bell}^\dagger=1_X
\eeq
and  of course 
\beq\label{traceunitarity2}
DeMeas_{Bell}^\dagger\circ DeMeas_{Bell}=1_{A\otimes A^*}\,.
\eeq
Let us verify that ${\cal M}_{Bell}$ is indeed a projector-valued spectrum.
Using eq.(\ref{traceunitarity}) we obtain $X$-idempotence:
\par\vspace{2mm}\par\noindent
\begin{minipage}[b]{1\linewidth}
\centering{\epsfig{figure=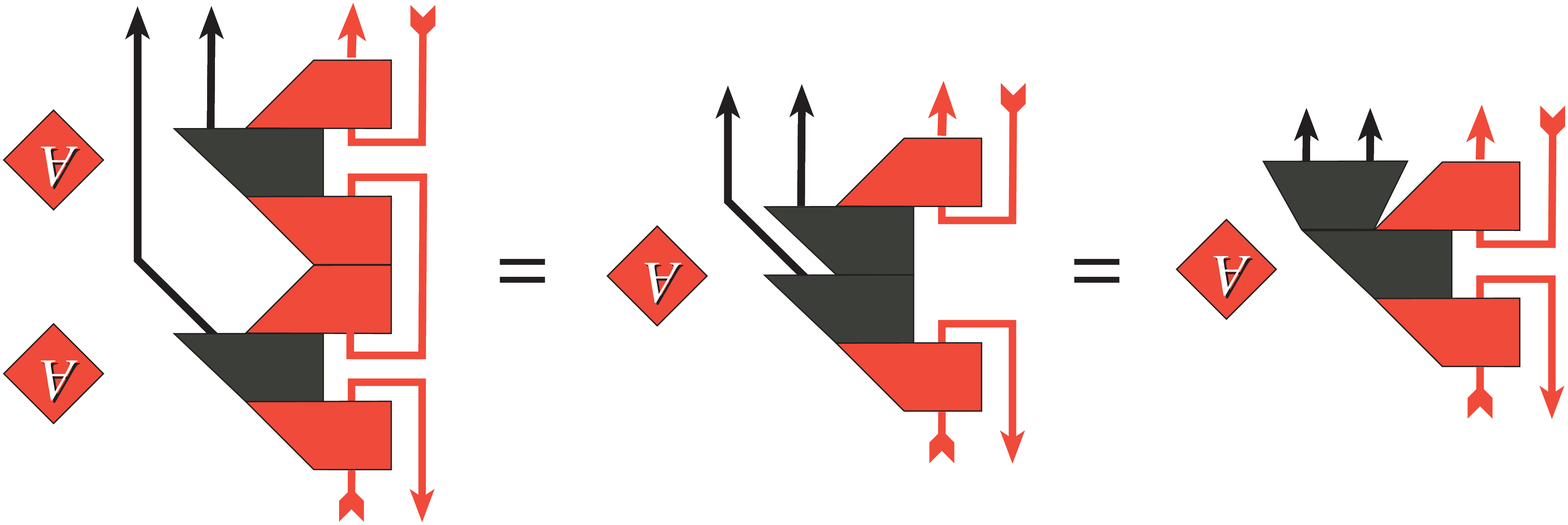,width=310pt}}        
\end{minipage}
\par\vspace{2mm}\par\noindent
and eq.(\ref{traceunitarity2})  assures $X$-completeness:
\par\vspace{2mm}\par\noindent 
\begin{minipage}[b]{1\linewidth}
\centering{\epsfig{figure=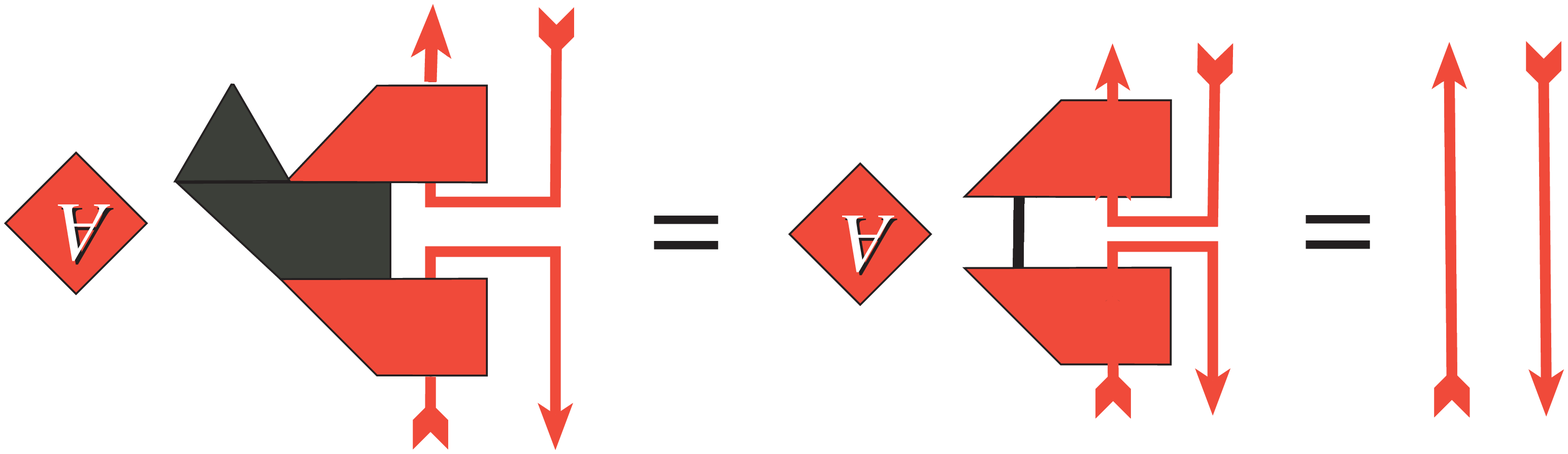,width=215pt}}        
\end{minipage}
\par\vspace{2mm}\par\noindent 
Finally, unitarity of $DeMeas_{Bell}$ yields $X\simeq A\otimes A^*$, so $DeMeas_{Bell}$ can be conceived as \em non-degenerate\em.  We normalize the \em Bell-states \em of type $A$ i.e.
\[
{1\over\sqrt{s_A}}\bullet\eta_A:\II\to A^*\otimes A\,.
\]

Now we will describe the teleportation protocol and prove its correctness. For simplicity we will not explicitly depict $\Gamma_X$ since it doesn't play an essential role in the topological manipulations of the picture.\footnote{A case where $\Gamma_X$ does play a crucial role is the proof of Naimark's theorem in  \cite{Paquette}.}    Here it is:
\par\vspace{2mm}\par\noindent  
\begin{minipage}[b]{1\linewidth}
\centering{\epsfig{figure=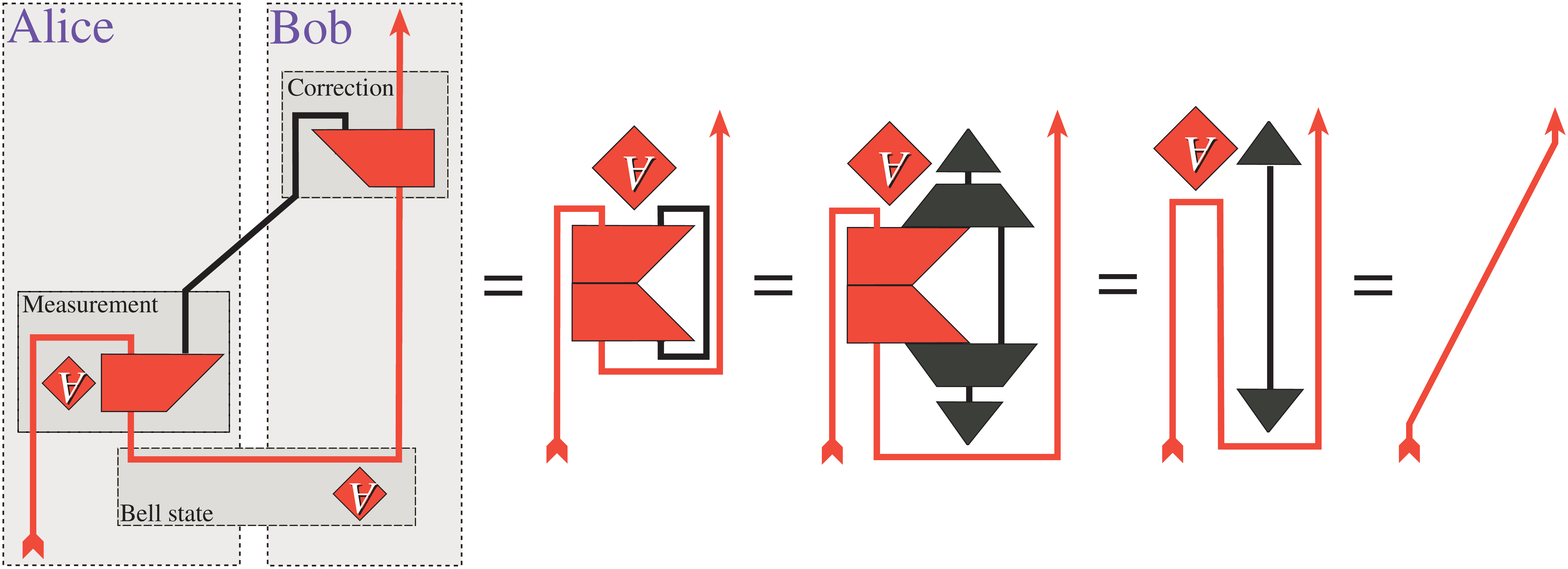,width=312pt}}        
\end{minipage}
\par\vspace{2mm}\par\noindent
%%%%%%%%%%%%%
The red box in {\tt Measurement} is a unitary morphism 
\[
\sigma_{X,A}\circ{\cal U}_*:A^*\to A^*\otimes X\,,
\]
which defines a demolition Bell-base measurement, the red box in {\tt Cor\-rection} is the unitary morphism 
\[
{\cal U}:A\to X\otimes A\,,
\]
and the bottom red box in the second picture obtained by `sliding' ${\cal U}$ along the red line is 
\[
{\cal U}^*\circ\sigma_{A,X}:A^*\otimes X\to A^*\,,
\]
the adjoint to $\sigma_{X,A}\circ{\cal U}_*$ --- note that the $\sigma$-isomorphisms are introduced to avoid crossing of lines.  Then we apply the decomposition $\eta:=\delta\circ\epsilon^\dagger$ which enables to use $X$-unitarity of ${\cal U}$. The reason why the black and the red scalar cancel out requires considering $\Gamma_X$ is part of the measurement --- we refer the reader to
\cite{CPP} for details.  We could \em copy \em the measurement outcome before \em consuming \em it:
%%%%%%%%%%%%%
\par\vspace{2mm}\par\noindent
\begin{minipage}[b]{1\linewidth}
\centering{\epsfig{figure=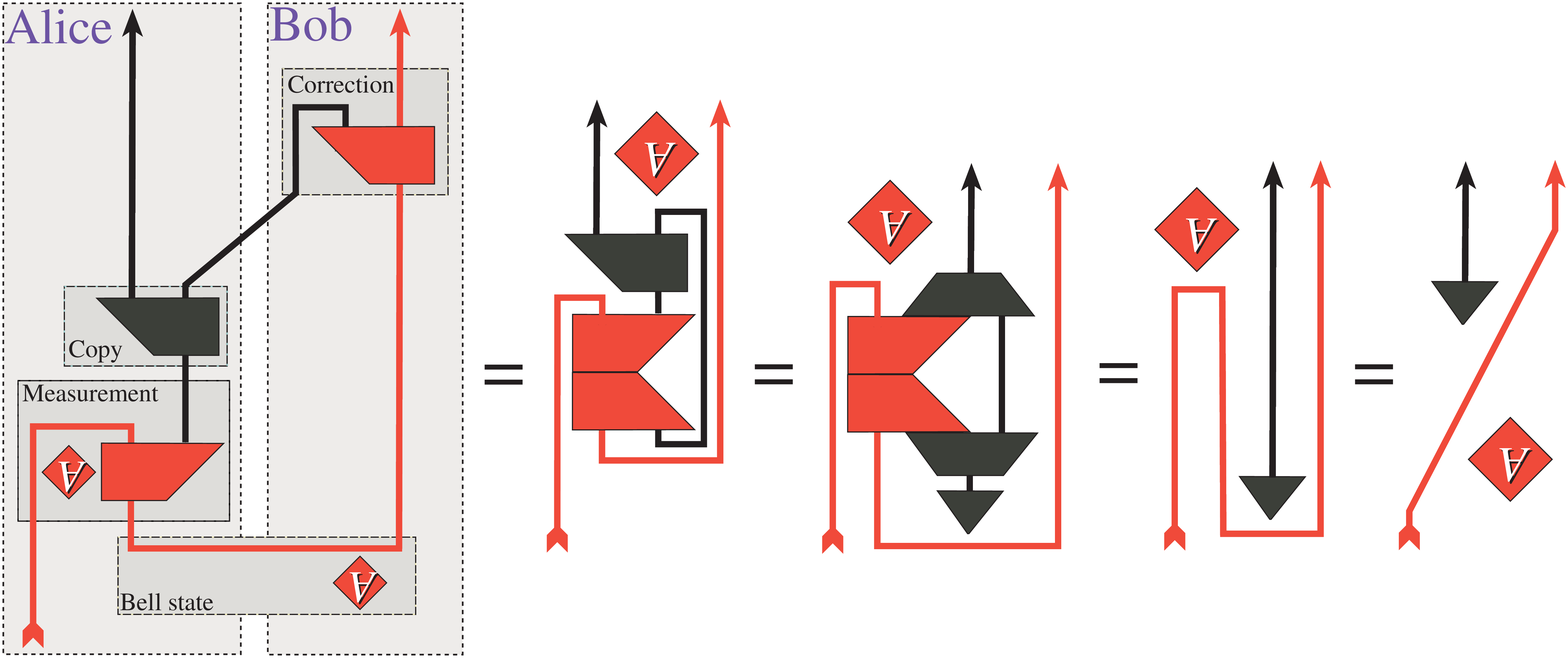,width=312pt}}    
\end{minipage}
\par\vspace{2mm}\par\noindent 
Note that  now we explicitly used $X$-self-adjointness. We can of course still choose to delete this data at a later stage using $\epsilon$, nicely illustrating \em resource-sensitivity\em.   If we wish to use the resulting available classical data for other purposes we possibly now might have to introduce $\Gamma_X$ explicitly.

Categorically we can fully specify this protocol as\footnote{The first specification of quantum teleportation as a commutative diagram, together with a purely categorical correctness proof,  is due to Abramsky and one of the authors \cite{AC1}.  However, their work relied heavily on the `unphysical' assumption of biproducts to establish this --- see \cite{deLL} for a discussion of this issue.}
\begin{scriptsize}
\begin{diagram}
A&\rTo^{(1_A\otimes\eta_A)\circ\rho_A}&A\otimes A^*\otimes A&\rTo^{DeMeas_{Bell}\otimes 1_A}&X\otimes A\\
\dTo^{\epsilon^\dagger\otimes1_A}&&&&\dIs\\
X\otimes A&\lTo^{1_X\otimes\left((\eta_X\otimes1_A)\circ(1_X\otimes{\cal U}^\dagger)\right)}&X\otimes X\otimes A&\lTo^{\delta\otimes 1_A}&X\otimes A
\end{diagram}
\end{scriptsize}
The morphism $\epsilon^\dagger\otimes1_A$ together with commutation of this diagram specifies the intended behavior, i.e.~teleporting a state of type $A$ with the creation of classical data as a bi-product,  while the other morphisms respectively are: (i) creation of a Bell-state $\eta_A$; (ii) a demolition Bell-base measurement $DeMeas_{Bell}$; (iii) copying of classical data using $\delta$; (iv) unitary correction using the $X$-adjoint to ${\cal U}$.
The above depicted graphical proof can be converted in an explicit category-theoretic one.

\section{Dense coding}\label{densecoding}

We can also give a similar description and proof of dense coding. 
\par\vspace{2mm}\par\noindent
\begin{minipage}[b]{1\linewidth}
\centering{\epsfig{figure=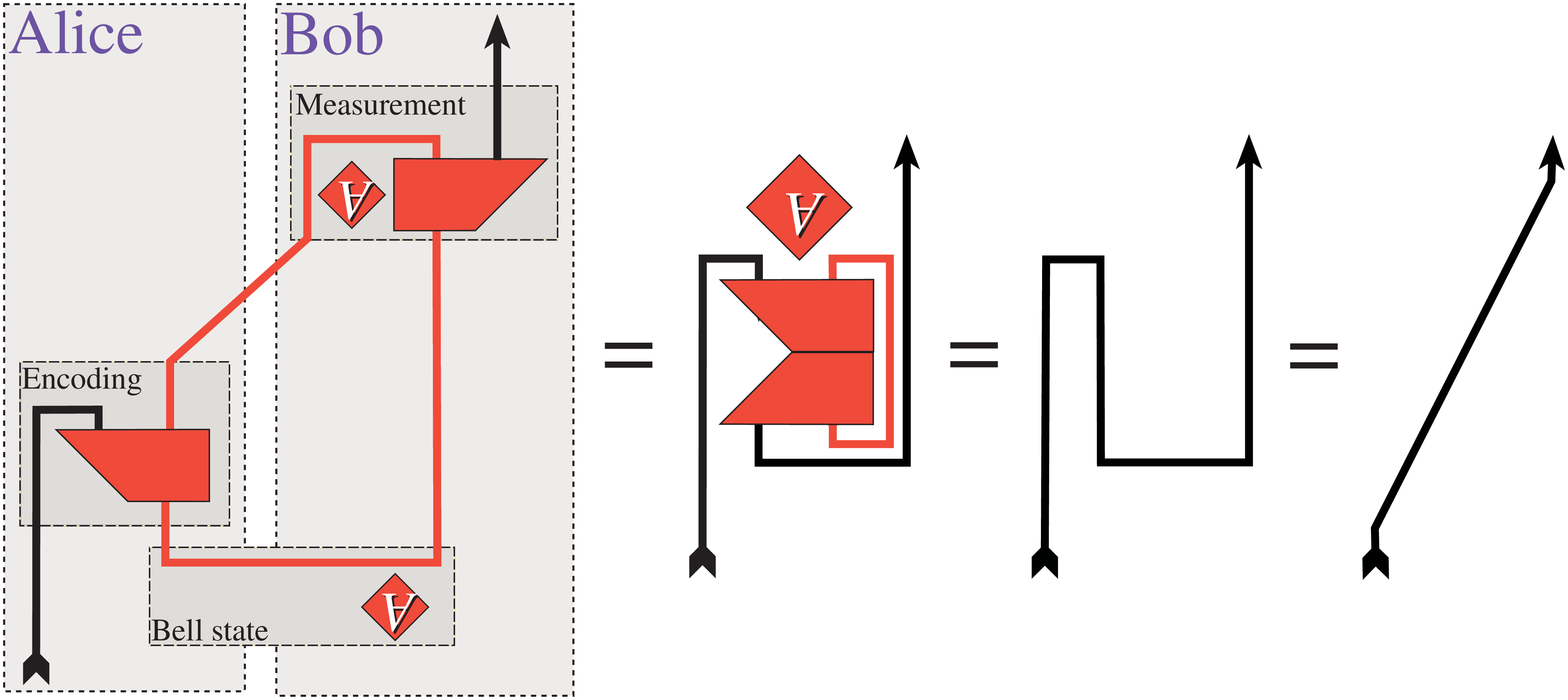,width=240pt}}   
\end{minipage}
\par\vspace{2mm}\par\noindent 
The remarks made above concerning  $\Gamma_X$ apply again here.
Note in particular that we rely on a very different property in this derivation than in the derivation of teleportation: here we use (one-sided) unitarity of $DeMeas_{Bell}$ while for teleportation we use $X$-unitarity of ${\cal U}$.  Hence it follows that teleportation and dense coding are not as closely related as one usually thinks: they are in fact \em axiomatically independent\em.

For a more systematic and more elaborate presentation of classical objects as the structure of classical data, together with several more quantum protocols, we refer the reader to \cite{CPP}.

\section*{Note and acknowledgements}

An earlier version of this chapter has been in circulation since November 2005 with as title \em Quantum measurements as coalgebras\em. This work was supported from the EPSRC grant EP/C500032/1 entitled {\em High-level methods in quantum computation and quantum information} and the NSF project 0209004 entitled {\em Coalgebraic methods for embedded and hybrid systems}.  Samson Abramsky, Dan Browne, Bill Edwards, Eric Oliver Paquette, Gordon Plotkin, Peter Selinger and Frank Valckenborgh provided useful comments. We used Paul Taylor's 2006 package to generate commutative diagrams.

\end{document}